\documentclass[journal]{IEEEtran}

\usepackage{amssymb}
\usepackage[cmex10]{amsmath}
\usepackage{stfloats}
\usepackage{graphicx}
\usepackage{subfigure}
\usepackage{tabularx}
\usepackage{verbatim}
\usepackage{url}
\usepackage{bm}
\usepackage{cite}
\usepackage{booktabs}
\usepackage[colorlinks]{hyperref}
\hypersetup{citecolor=blue}
\usepackage{balance}
\usepackage{amsmath}

\usepackage{algorithm}
\usepackage{algorithmic}
\usepackage{datetime}
\usepackage{epstopdf}

\usepackage{color}
\definecolor{myc1}{rgb}{0,0,0}

\usepackage[font=small,skip=2pt]{caption}

\usepackage{enumitem}
\setlist{nosep}

\begin{document}
\bstctlcite{IEEEexample:BSTcontrol}
\title{Fluid Antenna Array-Enabled Near-Field ISAC: \\Joint Port Position Optimization and \\Transmit Beamforming Design}

\author{Wangqi Shi, \IEEEmembership{Graduate Student Member, IEEE,} Zhentian Zhang, \IEEEmembership{Member, IEEE,} \\ Hao Jiang, \IEEEmembership{Senior Member, IEEE,} Zaichen Zhang,~\IEEEmembership{Senior Member,~IEEE}, \\
	and Kin-Fai Tong,~\IEEEmembership{Fellow,~IEEE}
\vspace{-11mm}

\thanks{}

\thanks{Wangqi Shi and Zaichen Zhang are with the National Mobile Communications Research Laboratory, Southeast University, Nanjing 210096, P. R. China (e-mails: shiwangqi@seu.edu.cn, zczhang@seu.edu.cn).}

\thanks{Zhentian Zhang and Kin-Fai Tong are with the Shenzhen Research Institute, Hong Kong Metropolitan University, Hong Kong, China (e-mails: ztzhang@hkmu.edu.hk, ktong@hkmu.edu.hk).}

\thanks{Hao Jiang is with the School of Cyber Science and Engineering, Southeast University, Nanjing 210096, P. R. China (e-mail: jiang.hao@seu.edu.cn).}

\thanks{Corresponding Author: Hao Jiang}
}

\maketitle

\begin{abstract}
Fluid antenna systems (FASs) have shown that activating a single port from a densely packed antenna array can yield spatial diversity comparable to that of tens of independent antennas, a property that has substantially improved communication reliability and efficiency. The fluid antenna array (FAA) extends this concept to multi-port configurations while avoiding complex multi-dimensional spatial correlation, and offers significant geometrical flexibility under a finite aperture and a limited port budget. These attributes motivate environment-adaptive array geometry for sensing in integrated sensing and communication (ISAC) systems. Although this flexibility enables adaptive array geometries that exploit near-field distance–angle information, improving multi-parameter target estimation under communication and deployment constraints remains challenging. In this paper, we propose a planar FAA-assisted near-field ISAC framework for multiuser communication and target sensing. Specifically, we derive the Fisher information matrix (FIM) and the corresponding Cramér–Rao bound (CRB) for the joint estimation of target range, zenith angle, azimuth angle, and radial velocity. Based on the derived CRB, we jointly optimize the FAA port positions and transmit beamforming to improve sensing accuracy while satisfying communication signal-to-interference-plus-noise ratio (SINR), transmit power, and physical deployment constraints. To solve the resulting non-convex problem, we develop a two-stage alternating optimization algorithm, in which transmit beamforming is optimized through semidefinite relaxation (SDR), while FAA port positions are successively optimized through discrete search and continuous refinement. Simulation results indicate that the proposed algorithm achieves favorable sensing performance while maintaining a reasonable computational cost.  
\end{abstract}

\IEEEpeerreviewmaketitle

\vspace{-2mm}
\section{INTRODUCTION}

\subsection{Background}
Integrated sensing and communication (ISAC) \cite{Liu-2022, Zhang-2026-ISAC,isac_tradeoff, Wei-2023} has emerged as a promising paradigm for next-generation wireless networks, improving spectral efficiency while reducing hardware costs. Beyond reliable data transmission, ISAC systems are expected to provide accurate estimates of target parameters, including position and velocity, to support environment-aware wireless applications \cite{Xu2025Elastic, Ahmadipour-2024, Demirhan-2023, ma_isac1,ma_isac2,ma_isac3}.  To meet the high accuracy demands of target sensing in ISAC systems, large-scale antenna arrays \cite{massive_MIMO1,massive_MIMO2,massive_MIMO3} are increasingly deployed. This aperture expansion extends the Rayleigh distance, thereby necessitating the consideration of near-field spherical-wave propagation \cite{Jiang-2023, Cui-2023}. Near-field spherical-wave models capture the dependence of array responses on both target range and angular directions, providing additional spatial information for target parameter estimation \cite{Liu-2023, Jiang-2025NearField}. However, traditional fixed-position arrays suffer from static geometric structures that make it difficult to fully release the potential diversities. 

Fortunately, fluid antenna systems (FASs) \cite{Wong2020,fas-twc-21,FBL3,FBL4} have demonstrated that activating a single port from an array of densely packed antennas can generate spatial diversity equivalent to that of dozens of independent antennas. Motivated by this capability, FAS-oriented designs have drastically improved communication reliability and efficiency \cite{FBL1,ML_WCNC,FBL2,yuanhui_wcl,zzt_sp1,FAS_channel_0_2,zzt_sp2,New-2023fluid,GBW_zzt,CSI1,Jiang-2025UAV}. Moreover, the fluid antenna array (FAA) \cite{Zhang-2026FAA,Zhang-2026-aperture,EM-FAA} provides a natural extension from single-port FAS to multi-port configurations while avoiding complex multi-dimensional correlation distributions. The FAA offers substantial geometrical flexibility in configuring multiple ports under finite aperture and limited port constraints, supporting applications such as beamforming/precoding, sidelobe suppression, and sensing \cite{wcsp_FAA1,wcsp_FAA2,FAA_XJY,wcsp_FAA3,wcsp_FAA4}. This reconfigurability facilitates environment-adaptive array geometry design and offers the potential to improve target parameter estimation accuracy.

\subsection{Related Works}
The estimation accuracy of near-field sensing systems has been extensively investigated, with the Cramér–Rao bound (CRB) widely adopted to characterize the fundamental limits of target parameter estimation \cite{Khamidullina-2021Bistatic, Wang-2024NearFieldCRB}. The authors in \cite{Hua-2024NearField3D} investigated the problem of three-dimensional near-field target localization under unknown reflection coefficients and in the presence of interference. Extending to ISAC systems, joint range-angle CRB minimization under communication-rate constraints was investigated in \cite{Wang-2023NearFieldISAC}, while \cite{Li-2024ISPAC} developed downlink and uplink sensing, positioning, and communication frameworks based on CRB optimization. The work in \cite{Zhang-2025DelayPhase} explored delay-phase precoding for CRB optimization under beam-squint effects. In addition, rate-splitting multiple access was considered for joint range and angle sensing in \cite{Zhou-2025NFISACRSMA}. Recent studies have also extended near-field sensing to cover more diverse target parameters, including angle-delay-Doppler estimation in hybrid near- and far-field scenarios \cite{Luo-2025AFDM} and joint estimation of three-dimensional position and radial velocity in broadband near-field ISAC \cite{Rahal-2026NearFieldISAC}. These studies collectively demonstrate the significance of array configuration and transmit signal design in near-field parameter estimation, thereby motivating further research on spatially reconfigurable antenna architectures.

Recent studies have explored FAS-assisted ISAC by exploiting spatial port reconfiguration to improve sensing and communication performance. Joint FAS port selection and transmit beamforming were investigated to reduce transmit power while meeting sensing and communication requirements \cite{Zou-2024FASISAC}, while the locations of fluid antennas and beamforming were jointly optimized under sensing beam pattern constraints in \cite{Zhou-2024FASISAC}. Other sensing-oriented studies considered radar signal-to-clutter-plus-noise ratio (SCNR) maximization under communication SINR constraints and CRB minimization through FAS port selection subject to communication requirements \cite{Ye-2025FASSCNR, Xu-2025FASCRB}. From a localization perspective, a Fisher information based framework was developed to derive positioning error bounds and guide FAS port selection \cite{Salem-2026FASLocalization}. In \cite{Zhang-2026FAA}, the authors investigated finite-aperture FAA design and derived a closed-form CRB to reveal the relationship between sensing accuracy and the geometric variance of port positions. Subsequently, continuous position optimization was carried out under the physical aperture constraint. Near-field FAS research further considered robust downlink beamforming and receive-port subset selection under imperfect channel state information \cite{Chen-2026NFFAS}. The authors in \cite{Guo-2026FA} investigated a fluid antenna enabled ISAC system and jointly optimized the FA positions, active beamforming, power allocation, and receive filters to minimize the CRB for target angle estimation while satisfying communication requirements. These studies highlight the potential of FAS reconfigurability and motivate further investigation of geometry-aware antenna design for near-field multi-parameter sensing under communication constraints.

Existing studies have established the importance of array geometry and transmit design in near-field parameter estimation and demonstrated the benefits of FAS port reconfiguration for sensing and communication performance. However, how to effectively exploit the spatial flexibility of planar FAAs for near-field multi-parameter target estimation while maintaining communication quality remains an important challenge. Under spherical-wave propagation, the sensing response depends on both the spatial parameters of targets and the active-port configuration, which allows the array geometry to influence the information available for parameter estimation. Meanwhile, reconfiguring the antenna ports also alters the communication channels, requiring sensing accuracy to be jointly considered with communication requirements and practical deployment constraints, including finite array aperture and minimum port spacing. Moreover, the strong coupling between FAA port configuration and transmit beamforming gives rise to a challenging non-convex optimization problem. Motivated by these considerations, this paper investigates a near-field ISAC system employing a planar FAA and develops a joint port position and transmit beamforming design to improve multi-parameter target estimation accuracy while meeting communication and physical deployment constraints.

\subsection{Main Contributions}

In this study, we propose a joint port position and transmit beamforming design for near-field ISAC systems employing planar FAAs to improve multi-parameter target estimation accuracy. The main contributions are summarized as follows:

\begin{itemize}
	\item We establish a near-field communication and sensing model employing a planar FAA with flexible port reconfiguration. To maintain a consistent maximum effective aperture across different port configurations, four corner ports are fixed, while the remaining active ports can be flexibly adjusted within a specified two-dimensional region. Based on spherical-wave propagation, the communication and sensing models are established to characterize the dependence of channel responses on the spatial configuration of active ports. Both line-of-sight (LoS) and non-line-of-sight (NLoS) propagation components are considered, with the NLoS-related sensing components modeled as aggregate multipath interference during target parameter estimation.
	
	\item We establish a sensing-oriented optimization model that incorporates communication requirements and practical FAA deployment constraints. The target estimation accuracy is characterized by the Fisher information matrix (FIM) and the corresponding CRB, with the normalized CRB adopted as the sensing performance indicator. The optimization model accounts for multiuser communication SINRs, total transmit power, finite array aperture, and minimum port spacing to ensure communication quality and physically feasible port configurations. In addition, a grid-based port density penalty is incorporated into the objective function to prevent excessive concentration of active ports. The resulting optimization problem captures the coupling among sensing accuracy, communication performance, and spatial port configuration.
	
	\item We develop a two-stage alternating optimization algorithm to address the non-convex coupling between FAA port configuration and transmit beamforming. For a given FAA configuration, the beamforming subproblem is reformulated using the Schur complement and handled through semidefinite relaxation (SDR). For fixed beamforming, a sequential discrete port position search is employed to update individual active ports within their feasible neighborhoods. The resulting discrete configuration is further refined in the continuous domain through a local search with progressively reduced step sizes. By alternately updating port positions and transmit beamforming in both stages, the proposed algorithm exploits the continuous spatial reconfigurability of the FAA while reducing the computational complexity of port position optimization.
	
	\item We conduct comprehensive numerical evaluations of the proposed FAA-assisted near-field ISAC framework from both algorithmic and sensing perspectives. The algorithmic aspects are first assessed through convergence and computational complexity analyses, together with benchmarks against exhaustive search and representative FAA schemes. The sensing characteristics are examined from aspects such as FAA reconfiguration, near-field angle focusing, and radial velocity sensing. Finally, we investigate the impacts of key system parameters such as transmit power, communication constraints, and the physical properties of the FAA on sensing performance and computational complexity.
\end{itemize}

The remainder of this paper is organized as follows. Section \ref{section_II} describes the planar FAA configuration and the corresponding near-field communication and sensing models. Section \ref{section_III} analyzes the sensing CRB and formulates the joint optimization problem under communication and deployment constraints. Section \ref{section_IV} details the proposed two-stage alternating optimization algorithm. Simulation results are presented in Section \ref{section_V}, followed by the conclusions in Section \ref{section_VI}.

\section{SYSTEM MODEL} \label{section_II}

\begin{figure}[!t]
	\centering
	\includegraphics[width=\columnwidth]{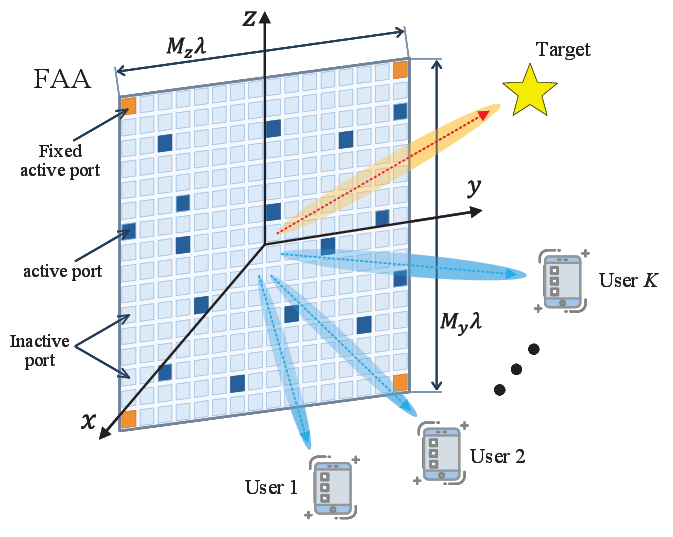}
	\caption{The 3D structural diagram of the proposed communication and sensing model.}\label{system_model}
	\vspace{-3mm}
\end{figure}

As shown in Fig. \ref{system_model}, we consider a near-field ISAC system in which a base station (BS) simultaneously serves multiple communication users and senses a target. The BS employs a finite-aperture planar fluid antenna array (FAA) for signal transmission, where multiple antenna ports are activated and their positions can be flexibly reconfigured within a prescribed two-dimensional aperture.

Port reconfiguration of the planar FAA may cause significant variations in the effective array aperture, which leads to inconsistent near-/far-field classification of the sensing target under different port configurations. To avoid this issue, we consider a finite-aperture planar FAA deployed at the BS \cite{Zhang-2026-aperture}. Specifically, the physical aperture size of the FAA is $\mathcal{A} = M_yM_z\lambda^2$, where $M_y\lambda$ and $M_z\lambda$ are the lengths of the FAA along the horizontal and vertical directions, respectively, with $\lambda$ being the wavelength. In the proposed FAA, $M$ antenna ports are simultaneously activated, with their 3D positions denoted by
\begin{align}
	\mathbf{P} = [\mathbf{p}_1, \mathbf{p}_2, ..., \mathbf{p}_m, ..., \mathbf{p}_M]^{\text{T}} \in \mathbb{R}^{M \times 3} ,
\end{align}
where $\mathbf{p}_m = (0, y_m, z_m)^{\text{T}}$ denotes the position of the $m$-th active port. To maintain the same maximum geometric aperture under different port configurations, i.e., a fixed near-field region, the four ports located at the vertices of the FAA are always activated:
\begin{align}\label{eq:fix_ports}
	\mathbf{p}_1 &= [0, -M_y\lambda/2, -M_z\lambda/2]^{\text{T}}, \mathbf{p}_2 = [0, M_y\lambda/2, -M_z\lambda/2]^{\text{T}}, \nonumber \\[0.125cm]
	\mathbf{p}_3 &= [0, -M_y\lambda/2, M_z\lambda/2]^{\text{T}}, \mathbf{p}_4 = [0, M_y\lambda/2, M_z\lambda/2]^{\text{T}} .
\end{align}
The remaining $M-4$ ports can move freely within the rectangular aperture.

In general, the maximum effective aperture is defined as the distance between the two farthest active ports. Thus, the effective aperture of the proposed FAA is $D_{\text{eff}} = \sqrt{(M_y\lambda)^2 + (M_z\lambda)^2}$. This paper adopts the Rayleigh distance as the criterion for dividing the near- and far-field regions, i.e.,
\begin{align}
	R_{\text{rayleigh}}= \frac{2 D_{\text{eff}}^2}{\lambda} = \frac{2 ((M_y\lambda)^2 + (M_z\lambda)^2)}{\lambda} .
\end{align}	
For example, for a $1 \text{m} \times 2 \text{m}$ FAA operating at $5$ GHz, the near-field region extends to approximately $166$ m, indicating that many practical sensing targets may fall within the near-field region and thus exhibit pronounced spherical-wave characteristics.

\subsection{Near-field Spherical Wave Response}

For any spatial point $\mathbf{q}$ located within the near-field region of the FAA, which may correspond to either a communication user or a sensing target, its position can be characterized by
\begin{align}
	\mathbf{q} (r, \theta, \phi) = r \begin{bmatrix}
		\sin\theta\cos\phi \\
		\sin\theta\sin\phi \\
		\cos\theta
	\end{bmatrix} ,
\end{align}	
where $r = \Vert \mathbf{q} \Vert_2$, and $\theta$ and $\phi$ denote the zenith and azimuth angles, respectively. The distance from the $m$-th active port in the FAA to point $\mathbf{q}$ can be expressed as
\begin{align}
	d_{\mathbf{p}_m, \mathbf{q}} &= \Vert \mathbf{q} (r, \theta, \phi) - \mathbf{p}_m \Vert_2  \nonumber \\[0.125cm]
	&= \sqrt{r^2 + y_m^2 + z_m^2 - 2r(y_m \sin\theta\sin\phi + z_m \cos \theta)} .
\end{align}	
Since the propagation phase related to $r$ is common across all activated FAA ports, this common phase term can be separated from the array response without affecting the relative phase relationships between the different ports. Here, we neglect the small amplitude variations among different ports, the near-field array response can be expressed as
\begin{align}\label{steering_vector}
	\mathbf{a} (r, \theta, \phi, \mathbf{P}) = \bigg[e^{-j \frac{2\pi}{\lambda}(d_{\mathbf{p}_1, \mathbf{q}} - r)}, ...,  e^{-j \frac{2\pi}{\lambda}(d_{\mathbf{p}_M, \mathbf{q}} - r)}\bigg]^{\text{T}} ,
\end{align}	
where $j = \sqrt{-1}$ denotes the imaginary unit. Unlike the conventional far-field array response, which depends only on the angular parameters, the near-field array response explicitly depends on both the target distance and the angular parameters. Therefore, the near-field response jointly captures the angular-domain and distance-domain characteristics of an arbitrary communication user or sensing target. Moreover, \eqref{steering_vector} characterizes the dependence of the near-field array response on the positions of the activated FAA ports, thereby providing a basis for optimizing the FAA port configuration.

\subsection{Near-field Communication and Sensing Model}

Based on the above near-field spherical-wave response, we next establish the communication and sensing signal models. Define the communication symbol vector of $K$ users at time instant $t$ as
\begin{align}
	\mathbf{c} (t) = [c_1 (t), c_2 (t), ..., c_K (t)]^{\text{T}} \in \mathbb{C}^{K \times 1} .
\end{align}	
Without loss of generality, we assume that the communication symbols of different users are mutually independent, i.e., $\mathbb{E}[\mathbf{c} (t) \mathbf{c}^{\text{H}} (t)] = \sigma^2_c \mathbf{I}_K$, where $\sigma^2_c$ denotes the average power. Similarly, the beamforming matrix is defined as
\begin{align}
	\mathbf{W} = [\mathbf{w}_1, \mathbf{w}_2, ..., \mathbf{w}_k, ..., \mathbf{w}_K] \in \mathbb{C}^{M \times K} ,
\end{align}	
where $\mathbf{w}_k$ denotes the beamforming vector corresponding to the $k$-th communication user. Therefore, the transmitted signal of the FAA can be expressed as
\begin{align}\label{eq:x_t}
	\mathbf{x} (t) =  \mathbf{W} \mathbf{c} (t) = \sum_{k=1}^{K} \mathbf{w}_k c_k (t) .
\end{align}	
It is worth emphasizing that the information symbol vector represents the data to be propagated to the users and thus is not treated as an optimization variable. The considered transmit design focuses on optimizing the spatial beamforming strategy rather than directly designing the communication waveform.

Based on the above derivation, the channel of the $k$-th communication user can be expressed as
\begin{align}
	\mathbf{h}_k (\mathbf{P}, t) = \sqrt{\frac{K_{\text{R}}}{K_{\text{R}}+1}} \mathbf{h}^{\text{LoS}}_k (\mathbf{P}, t) +  \sqrt{\frac{1}{K_{\text{R}}+1}} \mathbf{h}^{\text{NLoS}}_k (\mathbf{P}, t) ,
\end{align}	
where $K_{\text{R}}$ denotes the Rician factor. The terms $\mathbf{h}^{\text{LoS}}_k (\mathbf{P}, t)$ and $\mathbf{h}^{\text{NLoS}}_k (\mathbf{P}, t)$ denote the channel coefficients for the LoS and NLoS propagation links, respectively. Specifically, the LoS component is given by
\begin{align}
	\mathbf{h}^{\text{LoS}}_k (\mathbf{P}, t) = \sqrt{\beta^{\text{LoS}}_{c_k}} \mathbf{a} (r^{\text{LoS}}_{c_k}, \theta^{\text{LoS}}_{c_k}, \phi^{\text{LoS}}_{c_k}, \mathbf{P}) e^{j 2\pi f^{\text{LoS}}_{D} t} ,
\end{align}	
where $\beta^{\text{LoS}}_{c_k}$ denotes the path loss from the FAA to the $k$-th communication user for the LoS propagation link, $r^{\text{LoS}}_{c_k}$ is the propagation distance, and $\theta^{\text{LoS}}_{c_k}$ and $\phi^{\text{LoS}}_{c_k}$ represent the zenith and azimuth angles for the LoS propagation link, respectively. Moreover, $f^{\text{LoS}}_D = v_k \cos \vartheta^{\text{LoS}}_k / \lambda$ represents the Doppler frequency, where $v_k$ and $\vartheta^{\text{LoS}}_k$ denote the motion velocity and the angle between the motion direction and the LoS propagation link, respectively. For the NLoS propagation link, we assume that there are $L$ NLoS paths, yielding
\begin{align}
	\mathbf{h}^{\text{NLoS}}_k (\mathbf{P}, t) &= \sqrt{\beta^{\text{NLoS}}_{c_k}}  \sum^{L}_{\ell=1} \sqrt{\rho_{c_k,\ell}} e^{j\varphi_{c_k, \ell}}  \nonumber \\[0.125cm]
	&~ \times \mathbf{a} (r^{\text{NLoS}}_{c_k, \ell}, \theta^{\text{NLoS}}_{c_k, \ell}, \phi^{\text{NLoS}}_{c_k, \ell}, \mathbf{P}) e^{j 2\pi f^{\text{NLoS}}_{D, \ell} t} ,
\end{align}	
where $\beta^{\text{NLoS}}_{c_k}$ denotes the path loss from the FAA to the $k$-th communication user for the NLoS propagation link, $\rho_{c_k,\ell}$ is the normalized power ratio of the $\ell$-th ($\ell = 1, 2, ..., L$) NLoS path satisfying $\sum_{\ell=1}^{L} \rho_{c_k,\ell} =1$, and $\varphi_{c_k, \ell}$ denotes the scattering phase and the common propagation phase of the $\ell$-th NLoS path. The parameters $r^{\text{NLoS}}_{c_k, \ell}$, $\theta^{\text{NLoS}}_{c_k, \ell}$, and $\phi^{\text{NLoS}}_{c_k, \ell}$ represent the propagation distance and the zenith and azimuth angles corresponding to the $\ell$-th NLoS path, respectively. Moreover, $f^{\text{NLoS}}_{D, \ell} = v_k \cos \vartheta^{\text{NLoS}}_{k,\ell} / \lambda$ represents the Doppler frequency, where $\vartheta^{\text{NLoS}}_{k,\ell}$ denotes the angle between the motion direction and the $\ell$-th NLoS propagation link.

Combining the channels of all users into the channel matrix $\mathbf{H} (\mathbf{P}, t) = [\mathbf{h}_1 (\mathbf{P}, t), \mathbf{h}_2 (\mathbf{P}, t), ..., \mathbf{h}_K (\mathbf{P}, t)]^{\text{T}}$, the received signals of all communication users can be expressed as
\begin{align}\label{eq:signal_c}
	\mathbf{y} (t) = \mathbf{H} (\mathbf{P}, t) \mathbf{W} \mathbf{c} (t) + \mathbf{n}_c (t) ,
\end{align}	
where $\mathbf{n}_c (t) \sim \mathcal{CN}(0, \sigma_{n,c}^2 \mathbf{I}_K)$.

For the sensing functionality, the communication signal radiated by the FAA is also incident on the sensing target. Since the transmitted communication symbols and the beamforming matrix are known at the BS, the corresponding target echoes can be exploited to estimate the target parameters. However, the sensing echo undergoes two-way propagation from the FAA to the target and then back to the FAA, where both the forward and return links may experience either LoS or NLoS propagation. Consequently, a complete target-related sensing path can correspond to one of four propagation conditions: LoS–LoS, LoS–NLoS, NLoS–LoS, or NLoS–NLoS. To provide a unified representation, the propagation paths of the FAA-to-target and target-to-FAA links are modeled separately and then combined to construct the overall sensing channel, i.e.,
\begin{align}\label{sensing_H}
	\mathbf{H}_s (\mathbf{P}, t) &= \sum_{\ell_T=0}^{L_T} \sum_{\ell_R=0}^{L_R} \sqrt{\beta^s_{\ell_T, \ell_R} \rho^s_{\ell_T, \ell_R} } e^{j\varphi^s_{\ell_T, \ell_R} } \nonumber \\[0.125cm]
	&\,\,\, \times \mathbf{a} (r^s_{\ell_T}, \theta^s_{\ell_T}, \phi^s_{\ell_T}, \mathbf{P}) \mathbf{a}^{\text{T}} (r^s_{\ell_R}, \theta^s_{\ell_R}, \phi^s_{\ell_R}, \mathbf{P}) \nonumber \\[0.125cm]
	&\,\,\, \times e^{j2\pi f^s_{D, \ell_T, \ell_R} t} ,
\end{align}	
where $\ell_T = 1, 2, ..., L_T$ and $\ell_R = 1, 2, ..., L_R$ represent the $\ell_T$-th FAA-to-target path and $\ell_R$-th target-to-FAA path for the NLoS propagation links, respectively, while $\ell_T = 0$ or $\ell_R = 0$ indicates a LoS path. The parameter $\beta^s_{\ell_T, \ell_R}$ denotes the path loss, and $\rho^s_{\ell_T, \ell_R}$ is the normalized power ratio satisfying $\sum_{\ell_T=0}^{L_T} \sum_{\ell_R=0}^{L_R} \rho^s_{\ell_T, \ell_R} = 1$. The term $e^{j\varphi^s_{\ell_T, \ell_R}}$ denotes the scattering phase and the common propagation phase. The parameters $r^s_{\ell_T}$, $\theta^s_{\ell_T}$, $\phi^s_{\ell_T}$, $r^s_{\ell_R}$, $\theta^s_{\ell_R}$, and $\phi^s_{\ell_R}$ denote the distance and angle parameters of the forward and return paths, respectively. Moreover, $f^s_{D, \ell_T, \ell_R} = \frac{v_{\text{tar}}}{\lambda} (\cos \vartheta_{\ell_T} + \cos \vartheta_{\ell_R} )$ represents the Doppler frequency, where $v_{\text{tar}}$ is the motion velocity of the target, and $\vartheta_{\ell_T}$ and $\vartheta_{\ell_R}$ denote the angles between the target motion direction and the propagation directions of the $\ell_T$-th forward path and the $\ell_R$-th return path, respectively.

The target echo signal acquired by the sensing receive array can be expressed as
\begin{align}\label{eq:r_t}
	\mathbf{r} (t) = \mathbf{H}_s (\mathbf{P}, t) \mathbf{W} \mathbf{c} (t) + \mathbf{n}_s (t) ,
\end{align}	
where $\mathbf{n}_s (t) \sim \mathcal{CN}(0, \sigma_{n,s}^2 \mathbf{I}_M)$.

\section{CRB ANALYSIS AND PROBLEM FORMULATION} \label{section_III}

In this section, we formulate the joint FAA port position and transmit beamforming optimization problem for the considered near-field ISAC system. We first introduce the communication and deployment constraints, then derive the CRB for target parameter estimation and establish the corresponding optimization problem.

\subsection{Communication and Deployment Constraints}
Based on the communication signal model in \eqref{eq:signal_c}, the received signal at the $k$-th user can be expressed as
\begin{align}
	y_k (t) = \mathbf{h}_k^{\text{T}} ({\mathbf{P}, t}) \mathbf{w}_k \mathbf{c}_k (t) + \sum_{\substack{i=1\\ i\neq k}}^{K} \mathbf{h}_k^{\text{T}} ({\mathbf{P}, t}) \mathbf{w}_i \mathbf{c}_i (t) +n_{c,k} (t) ,
\end{align}	
where the first term denotes the desired signal, the second term represents the multiuser interference, and $n_{c,k} (t) \sim \mathcal{CN}(0, \sigma_{n,c}^2)$ denotes the additive noise. For notational simplicity, the time index $t$ is omitted in the following communication constraints when the channel realization at a given time instant is considered. Accordingly, the signal-to-interference-plus-noise ratio (SINR) of the $k$-th communication user is given by
\begin{align}
	\gamma_k (\mathbf{P,W}) = \frac{\sigma^2_c \vert \mathbf{h}_k^{\text{T}} (\mathbf{P}) \mathbf{w}_k \vert^2} { \sigma_c^2 \sum_{i=1,i\neq k}^{K} \vert \mathbf{h}_k^{\text{T}} (\mathbf{P}) \mathbf{w}_i \vert^2 + \sigma_{n,c}^2 }  .
\end{align}	
Since the objective of this paper is to optimize the sensing performance while satisfying the communication quality of service, a minimum SINR threshold is imposed on each user, that is
\begin{align}
	\gamma_k (\mathbf{P,W}) \geq \gamma_{\text{min}}, ~ k = 1, 2, ..., K ,
\end{align}	
where $\gamma_{\text{min}}$ denotes the prescribed minimum SINR threshold. This constraint ensures that the optimization of the FAA port positions and transmit beamforming for sensing enhancement does not result in an unacceptable degradation of the communication performance.

In addition to the communication requirement, the total transmit power of the FAA is subject to a maximum power budget. According to \eqref{eq:x_t}, the average transmit power can be obtained as
\begin{align}
	P_{\text{tx}} &= \mathbb{E}\left[\|\mathbf{x}(t)\|_2^2\right] \nonumber \\[0.125cm]
	&= \mathbb{E}\left[\mathbf{c}^H(t)\mathbf{W}^H\mathbf{W}\mathbf{c}(t)\right] \nonumber \\[0.125cm]
	&= \sigma_c^2 \operatorname{Tr}\left(\mathbf{W}\mathbf{W}^H\right) .
\end{align}	
By absorbing the symbol variance $\sigma_c^2$ into the transmit power budget, the power constraint can be equivalently written as
\begin{align}
	\operatorname{Tr}\left(\mathbf{W}\mathbf{W}^H\right) \leq P_{\text{max}} ,
\end{align}	
where $P_{\text{max}}$ denotes the maximum available transmit power at the BS.

Furthermore, the reconfiguration of the active FAA ports is restricted by the physical aperture and the minimum inter-port spacing. As introduced in Section II-A, the four ports located at the vertices of the FAA are fixed to maintain an invariant maximum geometric aperture under different port configurations, whereas the remaining $M-4$ ports can be flexibly repositioned within the prescribed rectangular aperture. Specifically, for $m = 5, ..., M$, we have 
\begin{align}
	-\frac{M_y\lambda}{2} &\leq y_m \leq \frac{M_y\lambda}{2}, & -\frac{M_z\lambda}{2} &\leq z_m \leq \frac{M_z\lambda}{2}.
\end{align}	
The minimum distance between different active ports is also restricted to avoid physical overlap, that is
\begin{align}
	\Vert \mathbf{p}_i - \mathbf{p}_j \Vert_2 \geq d_{\text{min}}, ~ \forall i \neq j ,
\end{align}
where $d_{\text{min}}$ denotes the minimum allowable spacing between any two active antenna ports, which ensures a physically feasible array configuration. The minimum inter-port spacing constraint only prevents physical overlap between active ports. However, it does not exclude the active ports from concentrating in a small local region of the FAA aperture. To discourage excessively clustered configurations while preserving the capability for continuous port reconfiguration of the FAA, we introduce a grid-based port density penalty. Specifically, the rectangular FAA aperture is partitioned into $G$ non-overlapping equal-area rectangular cells. For the $g$-th ($g=1,2,...,G$) grid cell, define the number of active ports located within it as
\begin{align}
	n_g(\mathbf{P})	&= \sum_{m=5}^{M} \mathbb{I}\left\{(y_m, z_m) \in \mathcal{G}_g\right\} ,
\end{align}
where $\mathbb{I}\{\cdot\}$ denotes the indicator function and $\mathcal{G}_g$ is the $g$-th grid cell. Based on the local occupancy, the following grid-based clustering penalty can be expressed as
\begin{align}
	\Phi_{\mathrm{grid}}(\mathbf{P}) &= \frac{1}{G} \sum_{g=1}^{G} \left[n_g(\mathbf{P}) - \frac{M-4}{G} \right]_{+}^2 ,
\end{align}
where $[x]_+ \triangleq \max\{x,0\}$. The proposed metric imposes no penalty when the number of active ports within a grid cell does not exceed the nominal occupancy. In contrast, when multiple ports become excessively concentrated within a local region, the penalty increases with the degree of local overcrowding. For convenience, we collect all feasible FAA port configurations into the set
\begin{align}
	\mathcal{P}_{\mathrm{FAA}}
	\triangleq \Big\{ \mathbf{P}: & -\frac{M_y\lambda}{2} \leq y_m \leq \frac{M_y\lambda}{2}, \nonumber \\[0.125cm]
	& -\frac{M_z\lambda}{2} \leq z_m \leq \frac{M_z\lambda}{2} ,
	~~ m=5, ..., M, \nonumber \\[0.125cm]
	& \Vert \mathbf{p}_i - \mathbf{p}_j \Vert_2 \geq d_{\text{min}}, ~ \forall i \neq j , \nonumber \\[0.125cm]
	& \, \mathbf{p}_1,\mathbf{p}_2,\mathbf{p}_3,\mathbf{p}_4
	\text{ are fixed as in \eqref{eq:fix_ports}}
	\Big\}.
\end{align}
Accordingly, the physical feasibility constraint on the FAA port configuration can be compactly written as
\begin{align}
	\mathbf{P} \in \mathcal{P}_{\mathrm{FAA}} .
\end{align}
Beyond the above port configuration constraints, we also aim to avoid excessively clustered port configurations. Therefore, the spatial dispersion of the active ports will be incorporated as an additional scoring term in the subsequent optimization problem.

\subsection{CRB Analysis}

To characterize the fundamental estimation accuracy of the sensing target, we derive the CRB for the target state parameters. Although the sensing channel in \eqref{sensing_H} contains four types of propagation components, the geometric parameters associated with the NLoS paths are assumed unavailable. Hence, only the LoS–LoS component is explicitly exploited for target parameter estimation, while the remaining components are treated as aggregate multipath interference. Accordingly, the sensing channel is decomposed as
\begin{align}
	\mathbf{H}_s (\mathbf{P},t) = \mathbf{H}_{\textit{LL}} (\mathbf{P},t) + \mathbf{H}_{\text{MP}} (\mathbf{P},t), 
\end{align}
where $\mathbf{H}_{\text{MP}} (\mathbf{P},t)$ collects all the components in \eqref{sensing_H} except for the LoS–LoS component. For the monostatic LoS–LoS path considered in this paper, the sensing channel is expressed as
\begin{align}
	\mathbf{H}_{\textit{LL}} (\mathbf{P},t) = \alpha e^{j\frac{4\pi v_r}{\lambda} t} \mathbf{a} (r, \theta, \phi, \mathbf{P}) \mathbf{a}^{\text{T}} (r, \theta, \phi, \mathbf{P}), 
\end{align}
where $\alpha = \alpha_R + j\alpha_I$ denotes the equivalent complex reflection coefficient that absorbs the LoS–LoS path gain and phase in \eqref{sensing_H}, and $v_r$ denotes the radial velocity of the target, such that $f^s_{D,0,0} = 2v_r/\lambda$. The remaining NLoS-related components are not explicitly exploited for target state estimation and are therefore incorporated into the aggregate multipath interference. Specifically, we have
\begin{align}
	\mathbf{z}_{\text{MP}} (t) \triangleq \mathbf{H}_{\text{MP}} (\mathbf{P},t) \mathbf{x} (t) , 
\end{align}
which is modeled as a zero-mean proper complex Gaussian random vector, that is, $\mathbf{z}_{\text{MP}} (t) \sim \mathcal{CN}(0, \mathbf{R}_{\text{MP}})$. Here, $\mathbf{R}_{\text{MP}}$ is treated as a fixed covariance matrix during beamforming optimization. Combining the multipath interference with the sensing receiver noise, the effective disturbance can be expressed as
\begin{align}\label{eq:m_t}
	\mathbf{m} (t) = \mathbf{z}_{\text{MP}} (t) + \mathbf{n}_s (t) \sim \mathcal{CN}(0, \mathbf{R}_{e}), 
\end{align}
where $\mathbf{R}_{e} = \mathbf{R}_{\text{MP}} + \sigma_{n,s}^2 \mathbf{I}_M$. For notational simplicity, define
\begin{align}\label{eq:a}
	\mathbf{a} \triangleq \mathbf{a}(r, \theta, \phi, \mathbf{P}), ~~ \mathbf{A} \triangleq \mathbf{a} \mathbf{a}^{\text{T}} . 
\end{align}
By substituting \eqref{eq:m_t} and \eqref{eq:a} into the sensing echo model in \eqref{eq:r_t}, the observation model used for target parameter estimation can be expressed as 
\begin{align}
	\mathbf{r} (t) = \alpha e^{j 2\pi f^s_{D,0,0}} \mathbf{A} \mathbf{x} (t) + \mathbf{m} (t). 
\end{align}
Define $\bm{\mu} (t) = \alpha e^{j 2\pi f^s_{D,0,0}} \mathbf{A} \mathbf{x} (t)$ as the mean of $\mathbf{r} (t)$, we have $\mathbf{r} (t) \sim \mathcal{CN}(\bm{\mu} (t), \mathbf{R}_{e})$. The parameters of interest are the target range, zenith angle, azimuth angle, and radial velocity, collected as
\begin{align}
	\bm{\eta} =  [r, \theta, \phi, v_r]^{\text{T}} .
\end{align}
Since the complex reflection coefficient $\alpha$ is unknown, its real and imaginary parts are further treated as nuisance parameters. Therefore, the complete real-valued unknown parameter vector is defined as
\begin{align}
	\bm{\xi} =  [r, \theta, \phi, v_r, \alpha_R, \alpha_I]^{\text{T}} .
\end{align}

To construct the FIM, we first derive the partial derivatives of $\bm{\mu} (t)$ with respect to the unknown parameters. We have all the derivatives in \eqref{eq:partial} (at the top of next page).
\begin{figure*}[!t]
	\normalsize
	\begin{subequations}\label{eq:partial}
		\begin{align}
			\frac{\partial \bm{\mu}(t)}{\partial r}
			&= \alpha e^{j2\pi f^s_{D,0,0}t}
			\left[ \frac{\partial \mathbf{a}}{\partial r}\mathbf{a}^{\text{T}}
			+ \mathbf{a}\left(\frac{\partial \mathbf{a}}{\partial r}\right)^{\text{T}} \right]\mathbf{x}(t),
			\label{eq:partial_a}\\[0.125cm]
			\frac{\partial \bm{\mu}(t)}{\partial \theta}
			&= \alpha e^{j2\pi f^s_{D,0,0}t}
			\left[ \frac{\partial \mathbf{a}}{\partial \theta}\mathbf{a}^{\text{T}}
			+ \mathbf{a}\left(\frac{\partial \mathbf{a}}{\partial \theta}\right)^{\text{T}} \right]\mathbf{x}(t),
			\label{eq:partial_b}\\[0.125cm]
			\frac{\partial \bm{\mu}(t)}{\partial \phi}
			&= \alpha e^{j2\pi f^s_{D,0,0}t}
			\left[ \frac{\partial \mathbf{a}}{\partial \phi}\mathbf{a}^{\text{T}}
			+ \mathbf{a}\left(\frac{\partial \mathbf{a}}{\partial \phi}\right)^{\text{T}} \right]\mathbf{x}(t),
			\label{eq:partial_c}\\[0.125cm]
			\frac{\partial \bm{\mu}(t)}{\partial v_r}
			&= j\frac{4\pi t}{\lambda}\bm{\mu}(t),
			\label{eq:partial_d}\\[0.125cm]
			\frac{\partial \bm{\mu}(t)}{\partial \alpha_R}
			&= e^{j2\pi f_{D,0,0}^{s}t}\mathbf{A}\mathbf{x}(t),
			\label{eq:partial_e}\\[0.125cm]
			\frac{\partial \bm{\mu}(t)}{\partial \alpha_I}
			&= j e^{j2\pi f_{D,0,0}^{s}t}\mathbf{A}\mathbf{x}(t).
			\label{eq:partial_f}
		\end{align}
	\end{subequations}
	\hrulefill
	\vspace*{4pt}
\end{figure*}
Assume that $T$ sensing observations are collected, indexed by $t=1,2,\ldots,T$. The observations are mutually independent, and the effective interference covariance matrix $\mathbf{R}_e$ is independent of the unknown parameter vector $\bm{\xi}$. Therefore, the $(i,j)$-th element of the FIM is given by the Slepian–Bangs formula as
\begin{align}\label{eq:FIM}
	[\mathbf{J}]_{i,j} &= 2\operatorname{Re} \left\{ \sum_{t=1}^{T} \left(	\frac{\partial \boldsymbol{\mu}(t)}	{\partial \xi_i} \right)^{\text{H}} \mathbf{R}_{e}^{-1} \left(	\frac{\partial \boldsymbol{\mu}(t)}	{\partial \xi_j} \right) \right\} ,
\end{align}
where $\xi_i$ denotes the $i$-th element of $\bm{\xi}$, and $\mathbf{J} \in \mathbb{R}^{6 \times 6}$ is the FIM associated with the complete unknown parameter vector. Since $\alpha_R$ and $\alpha_I$ are nuisance parameters, the FIM is partitioned as
\begin{align}
	\mathbf{J} =
	\begin{bmatrix}
		\mathbf{J}_{\eta\eta} & \mathbf{J}_{\eta\alpha} \\
		\mathbf{J}_{\alpha\eta} & \mathbf{J}_{\alpha\alpha}
	\end{bmatrix} ,
\end{align}
where $\mathbf{J}_{\eta\eta} \in \mathbb{R}^{4 \times 4}$ corresponds to the target state vector $\bm{\eta}$, $\mathbf{J}_{\alpha\alpha} \in \mathbb{R}^{2 \times 2}$ corresponds to the nuisance parameters, and $\mathbf{J}_{\eta\alpha} = \mathbf{J}^{\text{T}}_{\alpha\eta}$ characterizes their coupling. By eliminating the nuisance parameters through the Schur complement, the effective FIM for $\bm{\eta}$ is obtained as
\begin{align}
	\mathbf{J_{\text{eff}}} = \mathbf{J_{\eta\eta}} - \mathbf{J_{\eta\alpha}} \mathbf{J^{-1}_{\alpha\alpha}}  \mathbf{J}_{\alpha\eta} .
\end{align}
Therefore, the CRB matrix for the target state parameters can be expressed as
\begin{align}\label{eq:CRB}
	\mathbf{C}_{\text{CRB}} (\mathbf{P, W}) = \mathbf{J^{-1}_{\text{eff}}} .
\end{align}
The diagonal entries of $\mathbf{C}_{\text{CRB}}$ denote the lower bounds on the estimation variances of $r, \theta, \phi$ and $v_r$, respectively. Since $\mathbf{a} (r, \theta, \phi, \mathbf{P})$ depends on the FAA port configuration $\mathbf{P}$, while $\mathbf{x} (t) = \mathbf{W} \mathbf{c} (t)$ is determined by the transmit beamforming matrix $\mathbf{W}$, the resulting CRB is jointly determined by $\mathbf{P}$ and $\mathbf{W}$.

\subsection{Problem Formulation}

Based on the above CRB analysis, we jointly optimize the FAA port configuration and the transmit beamforming matrix to enhance target sensing accuracy while guaranteeing the communication quality of service. 
Since the target state vector contains four physical parameters, simply adding their CRBs may lead to the optimization objective being dominated by parameters with large numerical values. To provide a dimensionless and balanced measure of the overall sensing accuracy, we introduce the diagonal normalization matrix
\begin{align}
	\mathbf{D} = \operatorname{diag}\left(r_{\mathrm{ref}}, \theta_{\mathrm{ref}}, \phi_{\mathrm{ref}}, v_{\mathrm{ref}}\right),
\end{align}
where $r_{\text{ref}}$, $\theta_{\text{ref}}$, $\phi_{\text{ref}}$, and $v_{\text{ref}}$ are fixed reference scales obtained from the CRBs under a nominal sensing configuration with a full-power matched-filter beamformer. The normalized CRB matrix is then defined as
\begin{align}
	\overline{\mathbf{C}}_{\mathrm{CRB}}(\mathbf{P},\mathbf{W}) = \mathbf{D}^{-1}\mathbf{C}_{\mathrm{CRB}}(\mathbf{P},\mathbf{W})\mathbf{D}^{-1}.
\end{align}
Accordingly, the overall sensing performance is characterized by $f_s (\mathbf{P},\mathbf{W}) = \operatorname{Tr}\left(\overline{\mathbf{C}}_{\mathrm{CRB}}(\mathbf{P},\mathbf{W})\right) $.
The joint optimization problem is formulated as
\begin{align}
	\text{(P1)}: \,\,\, &\min_{\mathbf{P},\mathbf{W}} f_s (\mathbf{P},\mathbf{W}) + \omega_g \Phi_{\mathrm{grid}}(\mathbf{P}) \nonumber \\[0.125cm]
	\text{s.t.} \,\,\, &\gamma_k(\mathbf{P},\mathbf{W}) \geq \gamma_{\min}, \, k = 1, ..., K , \nonumber \\[0.125cm]
	& \operatorname{Tr}(\mathbf{W}\mathbf{W}^{\text{H}}) \leq P_{\max}, \nonumber \\[0.125cm]
	& \, \mathbf{P} \in \mathcal{P}_{\text{FAA}} ,
\end{align}
where $\omega_g \geq 0$ denotes the weighting factor for the grid-based clustering penalty. Problem (P1) jointly optimizes the FAA port configuration $\mathbf{P}$ and the transmit beamforming matrix $\mathbf{W}$. Due to the nonlinear dependence of the near-field array response on $\mathbf{P}$ and the coupled dependence of the CRB and communication SINR on $\mathbf{P}$ and $\mathbf{W}$, problem (P1) is generally non-convex. Therefore, directly obtaining its globally optimal solution is challenging, which motivates the development of an efficient joint optimization algorithm in the following section.

\section{ALGORITHM DESIGN} \label{section_IV}

In this section, we develop a two-stage alternating optimization algorithm to address the non-convex problem formulated in Section III. The proposed algorithm alternates between transmit beamforming optimization and FAA port position optimization, and combines discrete port position search with continuous refinement.

\subsection{Beamforming Optimization for Given FAA Configuration}

For a given feasible FAA port configuration \(\mathbf{P} \in \mathcal{P}_{\text{FAA}}\), the near-field array responses and the corresponding communication channels are fixed. We therefore first optimize the transmit beamforming matrix \(\mathbf{W}\) to minimize the sensing CRB while satisfying the communication SINR requirements and the total transmit power constraint.
\begin{align}
	\text{(P2)}: \,\,\, &\min_{\mathbf{W}} f_s (\mathbf{P},\mathbf{W}) \nonumber \\[0.125cm]
	\text{s.t.} \,\,\, &\gamma_k(\mathbf{P},\mathbf{W}) \geq \gamma_{\min}, \, k = 1, \ldots, K , \nonumber \\[0.125cm]
	&\operatorname{Tr}(\mathbf{W}\mathbf{W}^{\text{H}}) \leq P_{\max} .
\end{align}
To facilitate the beamforming design, we introduce the transmit covariance matrix \(\mathbf{V}_k \triangleq \mathbf{w}_k \mathbf{w}_k^{\text{H}}\) for the \(k\)-th user, which satisfies \(\mathbf{V}_k \succeq 0\) and \(\text{rank}(\mathbf{V}_k) = 1\). The received signal power can then be expressed as
\begin{align}
	\vert \mathbf{h}_k^{\text{T}} (\mathbf{P}) \mathbf{w}_k \vert^2 &= \mathbf{w}^{\text{H}}_k \mathbf{h}^*_k (\mathbf{P}) \mathbf{h}^{\text{T}}_k (\mathbf{P}) \mathbf{w}_k \nonumber \\[0.125cm]
	&= \operatorname{Tr}(\mathbf{h}^*_k (\mathbf{P}) \mathbf{h}^{\text{T}}_k (\mathbf{P}) \mathbf{w}_k  \mathbf{w}^{\text{H}}_k) \nonumber \\[0.125cm]
	&= \operatorname{Tr}(\mathbf{h}^*_k (\mathbf{P}) \mathbf{h}^{\text{T}}_k (\mathbf{P}) \mathbf{V}_k) .
\end{align}
Accordingly, the communication SINR constraints are rewritten as
\begin{align}\label{eq:c_SINR}
	\operatorname{Tr} &(\mathbf{h}^*_k (\mathbf{P}) \mathbf{h}^{\text{T}}_k (\mathbf{P}) \mathbf{V}_k) \nonumber \\[0.125cm]
	&\geq \gamma_{\text{min}} \left(\sum_{i=1,i\neq k}^{K} \operatorname{Tr}(\mathbf{h}^*_k (\mathbf{P}) \mathbf{h}^{\text{T}}_k (\mathbf{P}) \mathbf{V}_i) + \frac{\sigma_{n,c}^2} {\sigma^2_c} \right) .
\end{align}
Similarly, since \(\mathbf{W} = [\mathbf{w}_1, \mathbf{w}_2, \ldots, \mathbf{w}_K]\), the total transmit power can be written as
\begin{align}
	\operatorname{Tr} (\mathbf{W} \mathbf{W}^{\text{H}}) = \sum_{k=1}^{K} \operatorname{Tr} (\mathbf{w}_k \mathbf{w}^{\text{H}}_k) = \sum_{k=1}^{K} \operatorname{Tr} (\mathbf{V}_k),
\end{align}
and the transmit power constraint is equivalently expressed as
\begin{align}\label{eq:c_power}
	\sum_{k=1}^{K} \operatorname{Tr} (\mathbf{V}_k) \leq P_{\max} .
\end{align}

Next, we reformulate the sensing CRB in terms of \(\{\mathbf{V}_k\}_{k=1}^{K}\). Although the instantaneous communication symbols are known at the BS, they vary across sensing snapshots. To obtain a beamforming design that is independent of a particular communication-data realization, we characterize the sensing performance using the FIM averaged over the communication symbols. Since the communication symbols satisfy \(\mathbb{E}[\mathbf{c}(t)\mathbf{c}^{H}(t)] = \sigma_c^2 \mathbf{I}_K\), the statistical transmit covariance matrix is given by
\begin{align}
	\mathbb{E} [\mathbf{x}(t) \mathbf{x}^{\text{H}}(t)] &= \mathbf{W} \mathbb{E} [\mathbf{c}(t) \mathbf{c}^{\text{H}}(t)]  \mathbf{W}^{\text{H}} = \sigma^2_c \sum_{k=1}^{K} \mathbf{V}_k .
\end{align}
To reveal the dependence of the sensing FIM on the transmit covariance matrix, note that the transmitted signal \(\mathbf{x} (t)\) is independent of the unknown target parameter vector \(\bm{\xi}\). For convenience, define
\begin{align}\label{eq:mu_xi}
	\frac{\partial \bm{\mu(t)}} {\partial \xi_i} = \frac{\partial\left(\alpha e^{j\frac{4\pi v_r}{\lambda}t}\mathbf{A}\right)} {\partial \bm{\xi}_i} \mathbf{x}(t) = \mathbf{D}_i(t)\mathbf{x}(t) .
\end{align}
Substituting \eqref{eq:mu_xi} into the Slepian-Bangs formula in \eqref{eq:FIM} yields
\begin{align}\label{eq:J}
	[\mathbf{J}]&_{i,j} = 2\operatorname{Re} \left\{ \sum_{t=1}^{T} \mathbf{x}^{\text{H}} (t) \mathbf{D}^{\text{H}}_i(t) \mathbf{R}^{-1}_e \mathbf{D}_j(t) \mathbf{x} (t) \right\} \nonumber \\[0.125cm]
	&= 2\operatorname{Re} \left\{ \sum_{t=1}^{T} \operatorname{Tr} \left( \mathbf{D}^{\text{H}}_i(t)  \mathbf{R}^{-1}_e \mathbf{D}_j(t) \mathbf{x} (t) \mathbf{x}^{\text{H}} (t) \right) \right\}.
\end{align}
It shows that the dependence of the sensing FIM on the transmitted communication symbols is completely characterized through \(\mathbf{x} (t) \mathbf{x}^{\text{H}} (t)\). Since the beamforming design relies on the statistical transmit covariance rather than a particular realization of the communication symbols, we define the corresponding statistical FIM \(\mathbf{J}_s\) as
\begin{align}
	[\mathbf{J}_s]&_{i,j} = \mathbb{E}\left[[\mathbf{J}]_{i,j}\right] \nonumber \\[0.125cm]
	&= 2\operatorname{Re} \left\{ \sum_{t=1}^{T} \operatorname{Tr} \left( \mathbf{D}^{\text{H}}_i(t)  \mathbf{R}^{-1}_e \mathbf{D}_j(t) \mathbb{E}[\mathbf{x} (t) \mathbf{x}^{\text{H}} (t)] \right) \right\}  \nonumber \\[0.125cm]
	&= 2 \sigma^2_c \operatorname{Re} \left\{ \sum_{t=1}^{T} \operatorname{Tr} \left( \mathbf{D}^{\text{H}}_i(t)  \mathbf{R}^{-1}_e \mathbf{D}_j(t) \left( \sum_{k=1}^{K} \mathbf{V}_k \right) \right) \right\} .
\end{align}
Thus, for a given FAA configuration \(\mathbf{P}\) and fixed \(\mathbf{R}_e\), the statistical sensing FIM \(\mathbf{J}_s\) is an affine function of the \(\{\mathbf{V}_k\}_{k=1}^{K}\). To facilitate the subsequent optimization, define the selection matrix \(\mathbf{E} = [\mathbf{I}_4, \mathbf{0}_{ 4\times 2}]^{\text{T}}\). Using the block matrix inversion identity, the CRB matrix for the target state parameters can be expressed as
\begin{align}
	\mathbf{C}^s_{\text{CRB}} (\mathbf{V}_k) = \mathbf{E}^{\text{T}} \mathbf{J}^{-1}_s \mathbf{E} .
\end{align}
To eliminate the matrix inverse, we introduce an auxiliary matrix \(\mathbf{U} \succeq 0\). By the Schur complement, for \(\mathbf{J}_s \succ 0\),
\begin{align}
	\mathbf{U} \succeq \mathbf{E}^{\text{T}} \mathbf{J}^{-1}_s \mathbf{E} \Longleftrightarrow  \begin{bmatrix} \mathbf{J}_s & \mathbf{E} \\ \mathbf{E}^{\text{T}} & \mathbf{U} \end{bmatrix} &\succeq 0.
\end{align}
Since \(\operatorname{Tr}(\mathbf{U})\) is minimized, the above inequality is tight at the optimum. Hence, the beamforming optimization problem can be reformulated as
\begin{align}
	\text{(P2.1)}: \,\,\, &\min_{\{\mathbf{V}_k\}_{k=1}^{K}, \mathbf{U}} \operatorname{Tr}(\mathbf{D}^{-2} \mathbf{U})  \nonumber \\[0.125cm]
	\text{s.t.} \,\,\,\, &\begin{bmatrix} \mathbf{J}_s & \mathbf{E} \\ \mathbf{E}^{\text{T}} & \mathbf{U} \end{bmatrix} \succeq 0 , \,\,  
	\eqref{eq:c_SINR}, \,\, \eqref{eq:c_power}, ~ \mathbf{V}_k \succeq 0, \nonumber \\[0.125cm]
	&\,\, \text{rank} (\mathbf{V}_k) = 1, ~ \forall k .
\end{align}
Because \(\mathbf{D}^{-2}\) is positive definite, minimizing \(\operatorname{Tr}(\mathbf{D}^{-2} \mathbf{U})\) makes the Schur-complement inequality tight at the optimum. To handle the remaining non-convexity, we apply SDR by dropping the rank-one constraints in (P2.1). The resulting relaxed problem is a convex semidefinite program (SDP) and can be efficiently solved using standard convex optimization tools. If the obtained solution satisfies \(\text{rank}(\mathbf{V}^*_k)=1\), the corresponding beamforming vectors are directly recovered via eigenvalue decomposition. Otherwise, Gaussian randomization is employed to generate feasible rank-one beamforming candidates, from which the candidate satisfying the communication SINR and transmit power constraints with the minimum sensing CRB is selected.

\subsection{Discrete Port Position Optimization}

Section IV-A efficiently optimizes the beamforming matrix for a given FAA configuration. However, the nonlinear dependence of the near-field array response on \(\mathbf{P}\) poses a challenge for optimizing the port positions directly in the continuous domain. To reduce the search complexity while obtaining a high-quality port configuration, we first restrict the active ports to a finite set of candidate positions and perform port position optimization in the discrete domain.

Specifically, let \(\Delta d\) denote the spatial discretization interval. The continuous aperture is discretized along the \(y\)- and \(z\)-directions, yielding the candidate grid
\begin{align}
	\mathcal{Q}	&= \Bigg\{\Big( 0,-\frac{M_y\lambda}{2} + i \Delta d, -\frac{M_z\lambda}{2} + j\Delta d \Big)^{\text{T}} : \nonumber \\[0.125cm]
	& i=0,1,\ldots,\left\lfloor \frac{M_x\lambda}{\Delta d} \right\rfloor, ~ j=0,1,\ldots,\left\lfloor\frac{M_z\lambda}{\Delta d} \right\rfloor \Bigg\} .
\end{align}
Here, \(\mathbf{q}_{i,j} \in \mathcal{Q}\) represents a candidate port position on the discrete aperture, and \(\Delta d\) controls the spatial resolution of the discrete position search. It is worth mentioning  that \(\Delta d\) is distinct from the minimum inter-port distance \(d_{\text{min}}\): \(\Delta d\) determines the resolution of the candidate positions, whereas \(d_{\text{min}}\) imposes the physical feasibility constraint on any pair of active ports. Accordingly, the discrete feasible set of FAA configurations is defined as
\begin{align}
	\mathcal{P}_{\mathrm{D}}
	\triangleq \Big\{ \mathbf{P}: \,\, & \mathbf{p}_m \in \mathcal{Q}, ~ m=5, \ldots, M, \nonumber \\[0.125cm]
	& \Vert \mathbf{p}_i - \mathbf{p}_j \Vert_2 \geq d_{\text{min}}, ~ \forall i \neq j , \nonumber \\[0.125cm]
	& \, \mathbf{p}_1,\mathbf{p}_2,\mathbf{p}_3,\mathbf{p}_4
	\text{ are fixed as in \eqref{eq:fix_ports}}
	\Big\}.
\end{align}
Since \(\mathcal{P}_{\mathrm{D}} \subseteq \mathcal{P}_{\mathrm{FAA}}\), the discrete optimization can be regarded as a restricted approximation of the original continuous port position optimization. For a given beamforming matrix \(\mathbf{W}\), the port position optimization problem in (P1) is therefore approximated in the discrete domain as
\begin{align}\label{eq:P3}
	\text{(P3)}: \,\,\, &\min_{\mathbf{P}} f_s (\mathbf{P},\mathbf{W}) + \omega_g \Phi_{\mathrm{grid}}(\mathbf{P}) \nonumber \\[0.125cm]
	\text{s.t.} \,\,\, &\gamma_k(\mathbf{P},\mathbf{W}) \geq \gamma_{\text{min}}, \, k = 1, \ldots, K , \nonumber \\[0.125cm]
	& \, \mathbf{P} \in \mathcal{P}_{\text{D}} .
\end{align}
Although (P3) restricts the active ports to a finite set of candidate positions, an exhaustive search over all feasible combinations of the \(M-4\) active ports still incurs prohibitive computational complexity. To address this issue, we develop a low-complexity sequential local search method, in which the active ports are updated one at a time while the remaining ports stay fixed.

To initialize the discrete array, the four fixed corner ports in \eqref{eq:fix_ports} are first selected. The remaining \(M-4\) ports are then placed successively on the candidate grid \(\mathcal{Q}\). At each step, candidate positions whose distances to any selected port are smaller than \(d_{\text{min}}\) are excluded, and the candidate maximizing the minimum distance to the currently selected ports is chosen. Specifically, if \(\mathcal{S}\) denotes the set of currently selected port positions, the next port is placed at
\begin{align}\label{eq:q_star}
	\mathbf{q}^{\star}
	= \underset{
		\substack{
			\mathbf{q} \in \mathcal{Q} \\
			\Vert\mathbf{q}-\mathbf{p}\Vert_2 \geq d_{\text{min}} \\
			\forall \mathbf{p} \in \mathcal{S}
		}
	}{\arg\max}
	\ \underset{\mathbf{p} \in \mathcal{S}}{\min}
	\Vert\mathbf{q}-\mathbf{p}\Vert_2.
\end{align}
The selected position is then added to \(\mathcal{S}\), and the procedure is repeated until all \(M\) ports are determined.

Starting from the initialized configuration, the active ports are subsequently updated for a given \(\mathbf{W}\). For each \(m=5,\ldots,M\), the \(m\)-th port examines its current grid point and the four immediately adjacent grid points along the two aperture dimensions. Candidates outside the aperture or violating the \(d_{\text{min}}\) constraint are directly excluded. Since changing a port position alters the communication channels, candidates that violate the SINR constraints in \eqref{eq:P3} are further excluded. Among the remaining candidates, the position yielding the minimum objective value of (P3) is selected to replace the current position of the \(m\)-th port.

Once a port is updated, its new position is retained for the subsequent port updates. The \(M-4\) active ports are thus examined sequentially on the basis of the latest FAA configuration, while a port remains unchanged if no neighboring candidate yields a feasible objective improvement. After all active ports have been examined, the procedure is repeated until no further improvement is achieved. The overall discrete port position optimization is summarized in Algorithm \ref{alg:discrete_position}.

It is worth noting that the initialization in Lines 1--6 of Algorithm \ref{alg:discrete_position} is performed only once to generate the initial discrete FAA configuration. In the subsequent alternating iterations, the discrete position optimization is performed directly from the latest port configuration without reinitialization.
\begin{algorithm}[t]
	\caption{Discrete Port Position Optimization}
	\label{alg:discrete_position}
	\begin{algorithmic}[1]
		\REQUIRE Fixed beamforming matrix \(\mathbf W\), candidate grid \(\mathcal Q\),
		minimum spacing \(d_{\text{min}}\), and SINR threshold \(\gamma_{\text{min}}\).
		\ENSURE Optimized discrete port configuration \(\mathbf P^{\star}_{\rm D}\).
		\STATE Initialize \(\mathcal{S}\) with the four fixed corner ports in \eqref{eq:fix_ports};
		\FOR{\(m=5\) to \(M\)}
		\STATE Select \(\mathbf q^{\star}\) from \(\mathcal Q\) according to \eqref{eq:q_star};
		\STATE \(\mathcal{S} \leftarrow \mathcal{S} \cup \{\mathbf q^{\star}\}\);
		\ENDFOR
		\STATE Construct \(\mathbf{P}^{(0)}\) from \(\mathcal{S}\) and set
		\(\mathbf{P} \leftarrow \mathbf{P}^{(0)}\);
		\REPEAT
		\STATE \(\mathbf P_{\rm old} \leftarrow \mathbf{P}\);
		\FOR{\(m=5\) to \(M\)}
		\STATE Examine \(\mathbf p_m\) and its four neighboring grid points;
		\STATE Discard candidates violating the \(d_{\text{min}}\) or the SINR constraints;
		\STATE Evaluate the objective of (P3) for the remaining candidates;
		\STATE Update \(\mathbf p_m\) to the candidate with the minimum objective value;
		\ENDFOR
		\UNTIL{\(\mathbf{P}=\mathbf{P}_{\text{old}}\)}
		\STATE \(\mathbf{P}^{\star}_{\text{D}}\leftarrow\mathbf{P}\);
	\end{algorithmic}
\end{algorithm}

\subsection{Continuous Refinement and Overall Alternating Optimization}

Although the discrete port position optimization in Section IV-B significantly reduces the complexity of the position search, the optimized port positions are restricted to the candidate grid $\mathcal{Q}$, and the resulting spatial resolution is therefore limited by $\Delta d$. To further exploit the continuous mobility of the FAA, we refine the port positions in the original continuous aperture after the discrete optimization. Specifically, the optimized discrete configuration serves as the initial point for the continuous refinement, so that only local position adjustments are required rather than a direct search over the entire continuous aperture. For a given beamforming matrix $\mathbf{W}$, the continuous port position refinement can be formulated as
\begin{align}\label{eq:P4}
	\text{(P4)}: \,\,\, &\min_{\mathbf{P}} f_s (\mathbf{P},\mathbf{W}) + \omega_g \Phi_{\mathrm{grid}}(\mathbf{P}) \nonumber \\[0.125cm]
	\text{s.t.} \,\,\, &\gamma_k(\mathbf{P},\mathbf{W}) \geq \gamma_{\text{min}}, \, k = 1, ..., K , \nonumber \\[0.125cm]
	& \, \mathbf{P} \in \mathcal{P}_{\text{FAA}} .
\end{align}
Unlike (P3), the active ports in (P4) are no longer restricted to $\mathcal{Q}$. Since the discrete optimization already provides a favorable port configuration, we employ sequential local refinement to efficiently improve the port positions in the continuous domain.

Specifically, for the $m$-th active port, let $\delta^{\ell}$ denote the refinement step at the $\ell$-th refinement level. Eight trial positions are generated around its current position along the horizontal, vertical, and diagonal directions, given by
\begin{align}\label{eq:trial_positions}
	\mathbf{p}_m + \delta \mathbf{d}, ~ \mathbf{d} \in \left\{(0, \pm 1,0)^\text{T}, (0, 0,\pm 1)^\text{T}, (0, \pm 1,\pm 1)^\text{T} \right\}.
\end{align}
Similar to the discrete position update in Section IV-B, trial positions that violate the aperture, minimum inter-port distance, or SINR constraints are first excluded. Among the remaining feasible trial positions, the one yielding the lowest objective value of (P4) is selected if it improves the current solution; otherwise, the current port position is retained. The active ports are updated sequentially for $m=5,\ldots,M$ based on the latest FAA configuration.

After all active ports have been examined, the sequential update is repeated with the current refinement step as long as further objective improvement can be achieved. Once a complete sweep yields no improvement, a more precise local search is performed. Specifically, the refinement step at the $\ell$-th level is defined as
\begin{align}\label{eq:delta_l}
	\delta^{\ell} = \frac{\delta^0}{2^\ell}, ~ \ell = 0,1,... \, ,
\end{align}
where $\delta^{0}=\Delta d/2$ is the initial refinement step. The refinement level is increased when no further improvement can be made at the current step size, and the procedure terminates when $\delta^{\ell} \leq \delta_{\text{min}}$, where $\delta_{\text{min}}$ is a prescribed minimum refinement step. Consequently, the search resolution is progressively increased around the optimized discrete configuration.

Building on the beamforming optimization in Section IV-A, the discrete port position optimization in Section IV-B, and the continuous refinement described above, we develop a two-stage alternating optimization framework for solving (P1). The first stage performs joint beamforming and port position optimization over the discrete candidate grid, while the second stage further refines the resulting solution in the original continuous aperture.
\begin{itemize}
	\item  The initialization in Lines 1–6 of Algorithm 1 is performed only once to generate the initial discrete FAA configuration. For the current port configuration, the beamforming matrix is first optimized according to Section IV-A. With the obtained beamforming matrix fixed, Algorithm \ref{alg:discrete_position} is then applied to update the port positions starting from the latest configuration without reinitialization. These two steps are alternately performed until convergence, yielding the discrete solution $(\mathbf P_{\text{D}}^{\star},\mathbf W_{\text{D}}^{\star})$.
\end{itemize}
\begin{itemize}
	\item  The grid restriction is subsequently removed, and $\mathbf P_{\rm D}^{\star}$ is employed as the initial port configuration for continuous refinement. For the current port configuration, the beamforming matrix is first optimized according to Section IV-A. With the obtained beamforming matrix fixed, the active ports are sequentially refined following \eqref{eq:trial_positions} and \eqref{eq:delta_l}. For each port position update, the refinement step starts from $\delta^0=\Delta d/2$ and is progressively reduced until $\delta^\ell\leq\delta_{\text{min}}$. The beamforming and continuous port position updates are alternately performed until convergence, yielding the final solution $(\mathbf P^\star,\mathbf W^\star)$.
\end{itemize}

\begin{algorithm}[t]
	\caption{Two-Stage Alternating Optimization for Joint Beamforming and Port Position Design}
	\label{alg:overall}
	\begin{algorithmic}[1]
		\REQUIRE Candidate grid $\mathcal{Q}$, initial refinement step $\delta^0$,
		minimum refinement step $\delta_{\text{min}}$, and system parameters;
		\ENSURE Optimized port configuration $\mathbf P^\star$ and
		beamforming matrix $\mathbf W^\star$;
		
		\STATE Generate the initial discrete port configuration $\mathbf P$
		using Lines 1-6 of Algorithm \ref{alg:discrete_position};
		
		\STATE \textbf{Stage I: Discrete alternating optimization}
		\REPEAT
		\STATE Optimize $\mathbf{W}$ for the current $\mathbf{P}$
		according to Section~IV-A;
		\STATE Optimize $\mathbf{P}$ for the current $\mathbf{W}$
		according to Algorithm~1 without reinitialization;
		\UNTIL{the discrete alternating optimization converges}
		\STATE $\mathbf{P}_{\text{D}}^{\star}\leftarrow\mathbf{P}$,
		$\mathbf{W}_{\text{D}}^{\star}\leftarrow\mathbf W$;
		
		\STATE \textbf{Stage II: Continuous alternating refinement}
		\STATE $\mathbf{P}\leftarrow\mathbf{P}_{\text{D}}^{\star}$;
		\REPEAT
		\STATE Optimize $\mathbf W$ for the current $\mathbf P$
		according to Section~IV-A;
		\STATE $\delta \leftarrow \delta^0$;
		\WHILE{$\delta > \delta_{\text{min}}$}
		\REPEAT
		\STATE $\mathbf{P}_{\text{old}} \leftarrow \mathbf{P}$;
		\STATE Sequentially refine the active ports
		$m=5,\ldots,M$ using \eqref{eq:trial_positions} with step $\delta$;
		\UNTIL{$\mathbf{P} = \mathbf{P}_{\text{old}}$}
		\STATE $\delta \leftarrow \delta/2$;
		\ENDWHILE
		\UNTIL{the continuous alternating refinement converges}
		
		\STATE $\mathbf{P}^{\star} \leftarrow \mathbf{P}$,
		$\mathbf{W}^\star\leftarrow\mathbf{W}$;
		\RETURN $\mathbf{P}^{\star}, \mathbf{W}^{\star}$
	\end{algorithmic}
\end{algorithm}
The complete two-stage alternating optimization procedure is summarized in Algorithm \ref{alg:overall}. The computational complexity of the proposed algorithm arises mainly from the beamforming and port position optimization. For the discrete position optimization, Algorithm \ref{alg:discrete_position} first generates an initial port configuration. Let $N_Q=\vert\mathcal{Q}\vert$ denote the number of candidate grid points. Since the positions of the $M-4$ active ports are determined sequentially, the initialization requires $\mathcal{O}\big((M-4)N_Q\big)$ operations. As this initialization is performed only once, it introduces no repeated overhead during the subsequent alternating optimization.

After initialization, each active port examines only its four neighboring grid points rather than the entire candidate grid. Hence, the position search complexity of one complete discrete sweep is $\mathcal{O}(M-4)$, compared with $\mathcal{O}\big((M-4)N_Q\big)$ for an exhaustive grid search. For the continuous refinement, each active port examines only eight local trial positions, resulting in the same linear search complexity for a given refinement step. Since the refinement step is successively halved from $\delta^0$ to $\delta_{\text{min}}$, the number of refinement levels grows logarithmically with $\delta^0 / \delta_{\text{min}}$. Accordingly, the position search complexity of the continuous refinement scales as $\mathcal{O}\big((M-4) \log_2 \frac{\delta^0}{\delta_{\text{min}}} \big)$.
\begin{table}[!t]
	\renewcommand{\arraystretch}{1.15}
	\caption{Default simulation parameters.}
	\label{tab_parameters}
	\centering
	\begin{tabular*}{\columnwidth}{@{}@{\extracolsep{\fill}}ll@{}}
		\hline
		Parameter & Value \\
		\hline
		Carrier frequency $f_c$ & $5$ GHz \\
		Wavelength $\lambda$ & $0.06$ m \\
		Aperture size & $0.5~\mathrm{m} \times 0.5~\mathrm{m}$ \\
		Number of active ports $M$ & $36$ \\
		Minimum inter-port spacing $d_{\min}$ & $0.2\lambda$ \\
		Number of users $K$ & $4$ \\
		Minimum SINR $\gamma_{\min}$ & $10$ dB \\
		Maximum transmit power $P_{\max}$ & $30$ dBm \\
		Target range $r$ & $10$ m \\
		Target zenith angle $\theta$ & $80^\circ$ \\
		Target azimuth angle $\phi$ & $30^\circ$ \\
		Radial velocity $v_r$ & $2$ m/s \\
		Rayleigh distance $R_{\mathrm{Rayleigh}}$ & $\approx16.67$ m \\
		\hline
	\end{tabular*}
	\vspace{-2mm}
\end{table}
\begin{figure}[!t]
	\centering
	\includegraphics[width=\columnwidth]{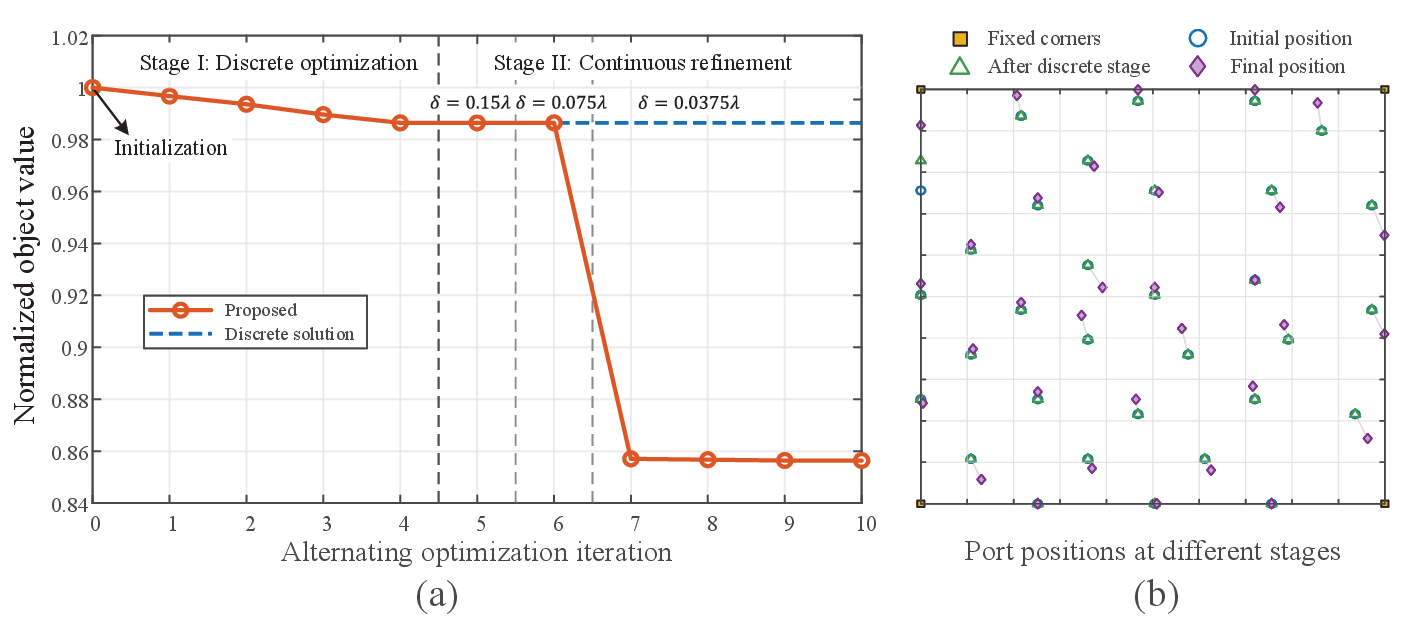}
	\caption{Convergence behavior and port position evolution of the proposed two-stage optimization algorithm: (a) convergence of the normalized objective value; (b) port positions at different optimization stages.}
	\label{fig_convergence}
	\vspace{-3mm}
\end{figure}
\begin{figure}[!t]
	\centering
	\includegraphics[width=\columnwidth]{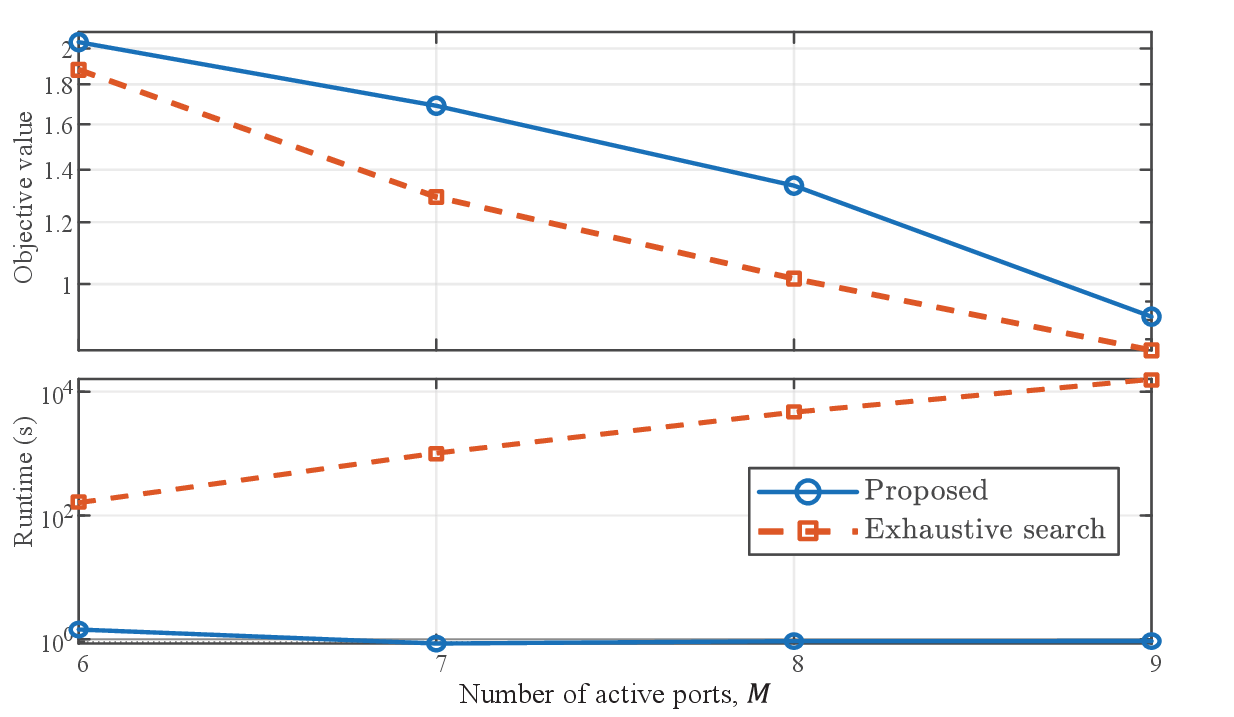}
	\caption{Performance and complexity comparison with exhaustive search.}
	\label{fig_benchmark}
	\vspace{-3mm}
\end{figure}

The above analysis characterizes the search complexity of the port position optimization. For each candidate position, the SINR constraints and the CRB objective need to be evaluated. In addition, the beamforming update in each alternating iteration requires solving the SDP formulated in Section IV-A, which constitutes the major computational cost of the overall algorithm. Nevertheless, after the one-time initialization, only a constant number of local candidate positions are examined for each active port. This avoids an exhaustive search over the entire discrete grid or continuous aperture, thereby maintaining a relatively low complexity for the position search.

\textit{Remark 1 (Motivation of the Two-Stage Design): The proposed two-stage design is motivated by the tradeoff between search complexity and sensing accuracy. Directly optimizing the port positions over the continuous aperture entails a prohibitively large search space, whereas restricting the optimization to a discrete grid inevitably limits the achievable positioning accuracy. Therefore, the discrete optimization is first employed to efficiently identify a favorable port configuration, which is subsequently refined in the continuous domain to recover the performance loss caused by spatial discretization.}

\textit{Remark 2 (Motivation of the Sequential Position Update): The sequential position update is adopted to avoid the high complexity of jointly searching the positions of all active ports. By optimizing one port at a time while keeping the others fixed, the high-dimensional position optimization is decomposed into a sequence of low-complexity local searches. Moreover, an update is accepted only when it yields an objective improvement while satisfying all constraints, ensuring a non-increasing objective value for fixed beamforming.}
\begin{figure*}[!t]
	\centering
	\includegraphics[width=\textwidth]{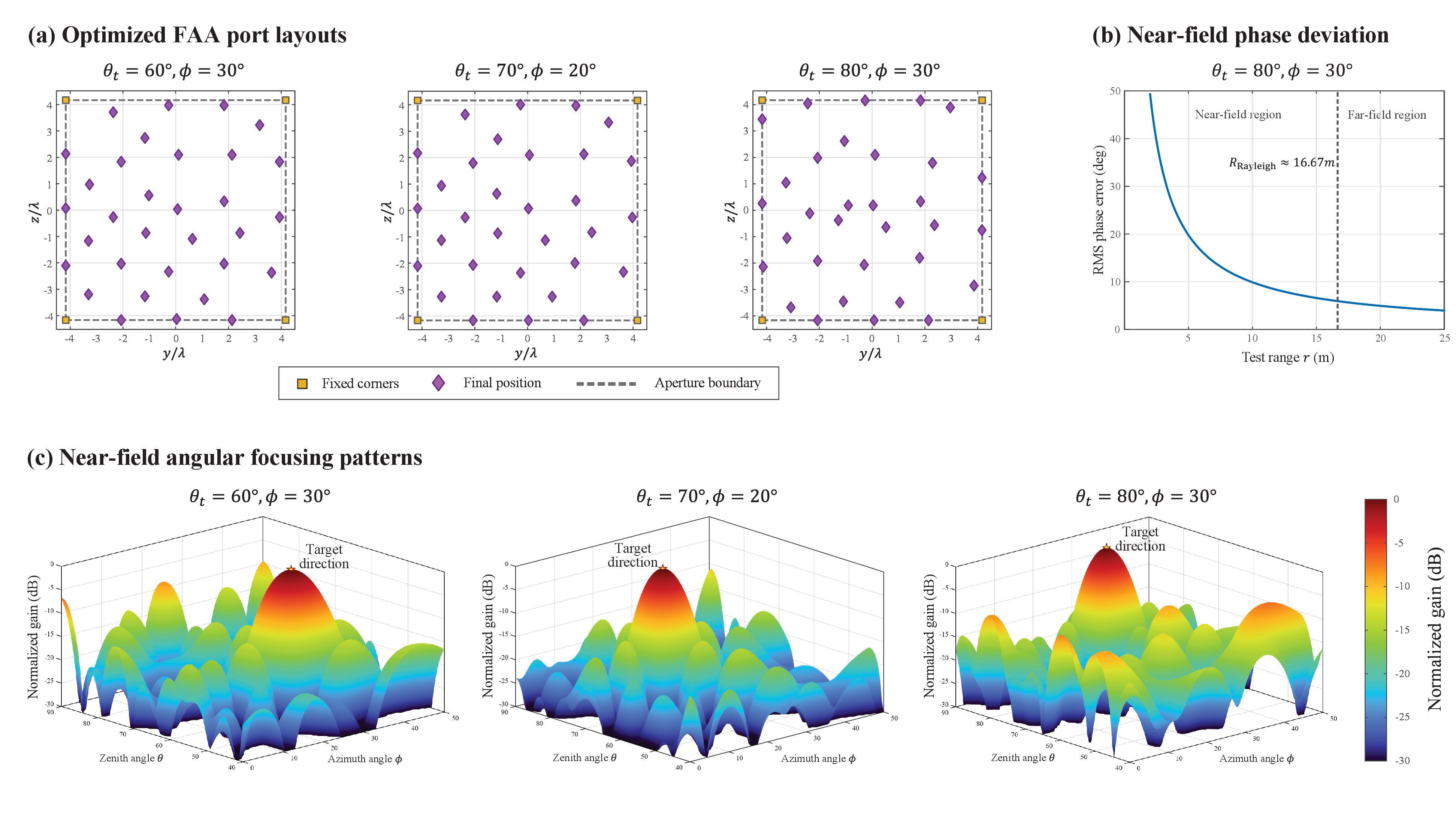}
	\caption{FAA reconfiguration and near-field focusing characteristics: (a) optimized FAA port layouts; (b) near-field phase deviation; (c) near-field angular focusing patterns.}
	\label{fig_focusing}
\end{figure*}
\begin{figure}[!t]
	\centering
	\includegraphics[width=\columnwidth]{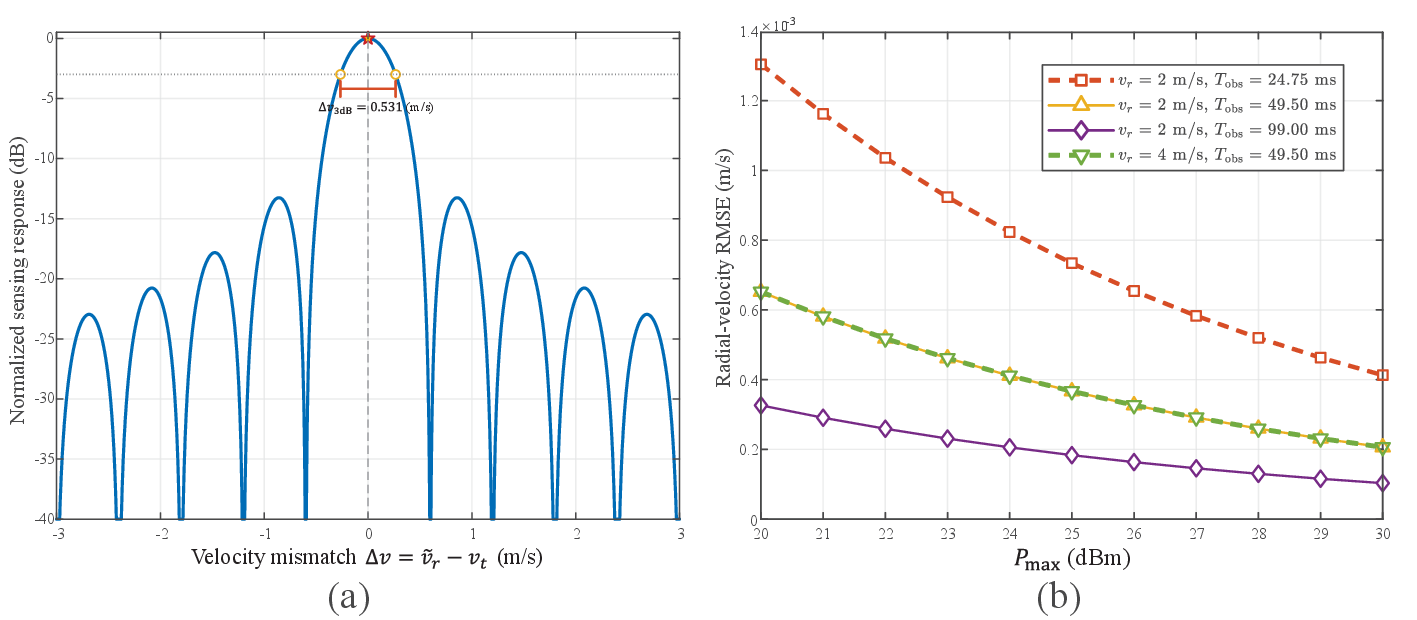}
	\caption{Radial velocity sensing characteristics and estimation performance: (a) normalized response versus velocity mismatch; (b) radial velocity root mean square error (RMSE) versus $P_{\text{max}}$. }
	\label{fig_velocity}
\end{figure}
\begin{figure}[!t]
	\centering
	\includegraphics[width=\columnwidth]{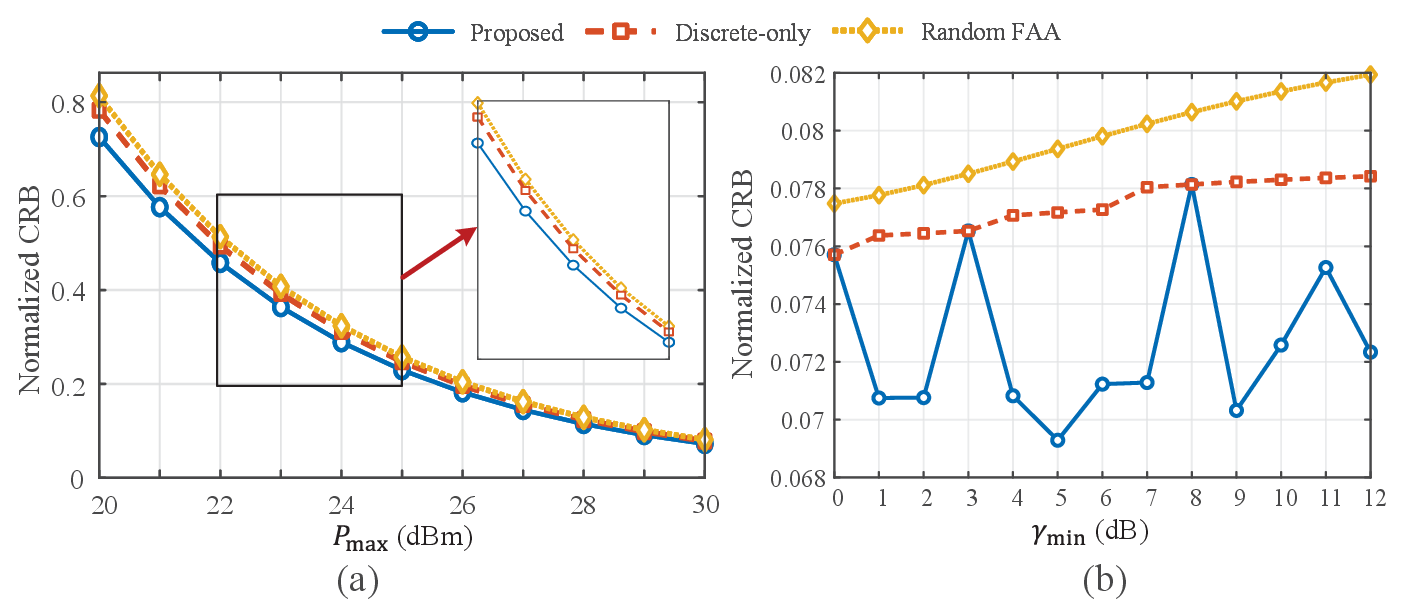}
	\caption{Performance comparison of normalized CRB under different FAA port configurations: (a) normalized CRB versus $P_{\text{max}}$; (b) normalized CRB versus $\gamma_{\text{min}}$.}
	\label{fig_random_poroposed}
\end{figure}
\begin{figure}[!t]
	\centering
	\includegraphics[width=\columnwidth]{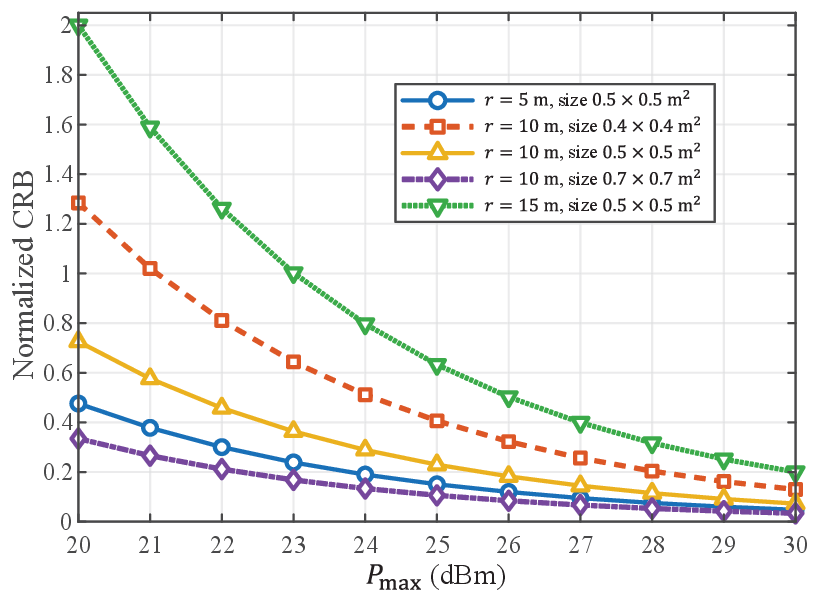}
	\caption{Impact of target range and aperture size on sensing performance.}
	\label{fig_r_W}
\end{figure}
\begin{figure}[!t]
	\centering
	\includegraphics[width=\columnwidth]{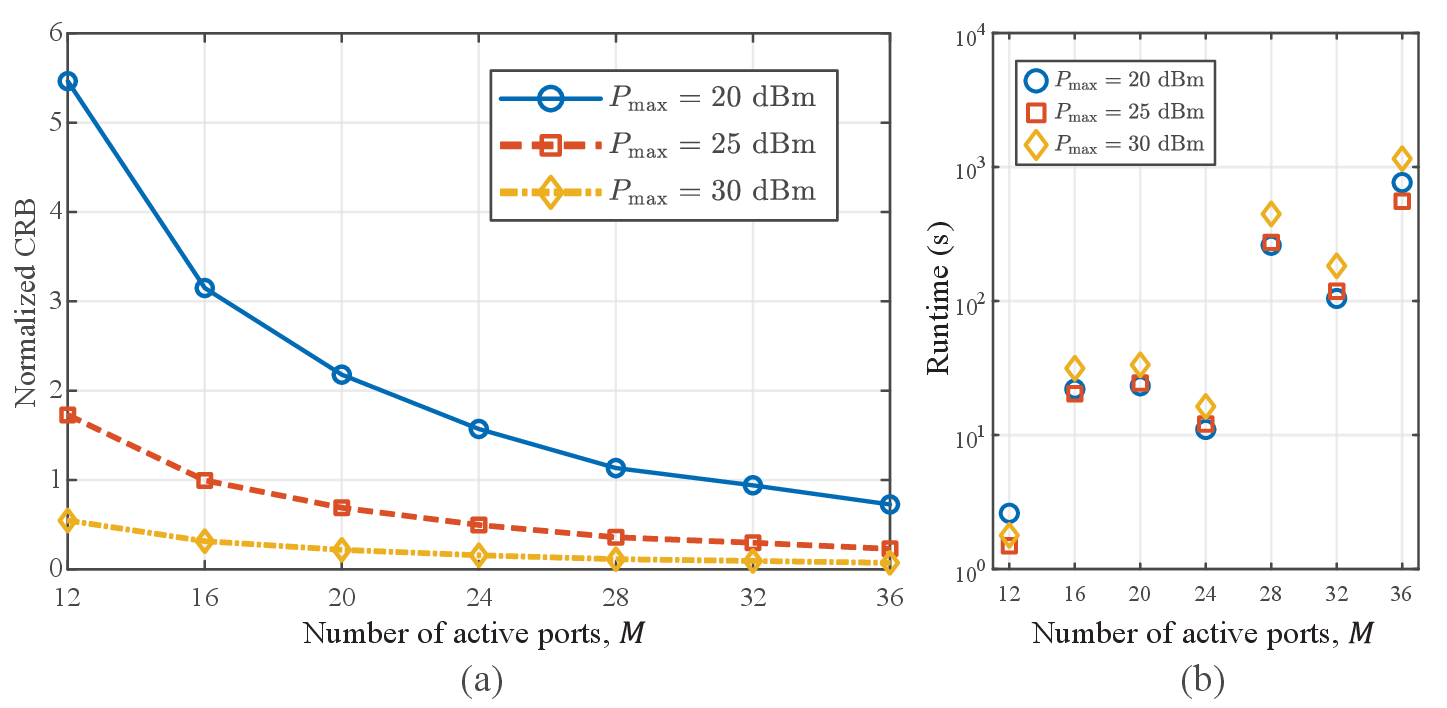}
	\caption{The relationship between sensing performance and computational complexity with respect to the number of active ports: (a) normalized CRB; (b) runtime.}
	\label{fig_P_max_runtime}
\end{figure}
\section{SIMULATION RESULTS} \label{section_V}

Unless otherwise specified, the default simulation parameters are summarized in Table \ref{tab_parameters}. With $R_{\mathrm{Rayleigh}}\approx16.67$ m and the default target at $r=10$ m, the target lies within the nominal near-field region.

Fig. \ref{fig_convergence} illustrates the convergence and port configuration evolution of the proposed algorithm. Stage I converges quickly to a stable feasible solution, yet further improvement remains available once the grid restriction is removed in Stage II, indicating that sensing performance is sensitive to port geometry. The discrete stage assigns ports over the aperture area, while continuous refinement introduces local displacements that overcome grid limitations and further improve the objective.

Fig. \ref{fig_benchmark} compares the proposed method with exhaustive search in a small-scale setting where the latter remains tractable. As $M$ increases, both methods achieve lower objective values, confirming that additional spatial degrees of freedom benefit the joint design. Although exhaustive search attains a slightly lower objective, it incurs substantially higher computational cost, demonstrating that the proposed algorithm offers a favorable performance-complexity tradeoff.

Fig. \ref{fig_focusing} shows that distinct target directions produce distinct optimized port configurations, indicating that the FAA geometry adapts to the specific sensing scenario rather than adopting a fixed layout. Phase deviation from the far-field planar-wave approximation becomes more pronounced at shorter distances, confirming that spherical-wave characteristics cannot be ignored in the near-field. The angular beampatterns in Fig. \ref{fig_focusing}(c) form main lobes around the corresponding target directions, demonstrating direction-dependent geometry adaptation and angular focusing.

Fig. \ref{fig_velocity}(a) shows a distinct main lobe around zero mismatch, confirming that the received signal carries sufficient Doppler information for velocity discrimination. In Fig. \ref{fig_velocity}(b), the RMSE decreases with increasing transmit power and with longer observation duration, the latter providing finer Doppler discrimination. Curves for different target velocities nearly overlap, indicating that accuracy is governed primarily by transmit power and observation duration rather than by the absolute radial velocity.

Fig. \ref{fig_random_poroposed}(a) compares the proposed algorithm with the discrete-only and random FAA baselines. The normalized CRB decreases with $P_{\text{max}}$ for all schemes, with the proposed algorithm consistently achieving the lowest CRB, followed by the discrete-only scheme and the random baseline. This confirms that geometry optimization and continuous refinement both contribute to sensing gain. Fig. \ref{fig_random_poroposed}(b) further shows that the proposed algorithm maintains a comparatively low CRB across the considered SINR range, sustaining good sensing performance under communication constraints.

Fig. \ref{fig_r_W} shows that the normalized CRB decreases with $P_{\text{max}}$ under all considered settings. For a fixed $0.5 \times 0.5 \text{m}^2$ aperture, sensing performance degrades as the target range increases from $5$ m to $15$ m, whereas enlarging the aperture from $0.4 \times 0.4 \text{m}^2$ to $0.7 \times 0.7 \text{m}^2$ at $r=10$ m reduces the CRB. These trends arise because shorter distances and larger apertures produce stronger spatial phase variations, increasing the sensitivity of the received signal to target parameters.

Fig. \ref{fig_P_max_runtime} shows that increasing $M$ reduces the normalized CRB under all transmit power budgets, although the improvement gradually weakens, indicating diminishing returns. Higher transmit power further lowers the CRB without altering the overall dependence on $M$. The computational cost generally increases significantly with $M$, with fluctuations attributable to differences in convergence behavior. The number of active ports should therefore be selected by balancing sensing accuracy against computational complexity.

\section{CONCLUSION} \label{section_VI}
We investigate FAA-boosted near-field multi-parameter target sensing in ISAC through joint port position and transmit beamforming design, and a two-stage alternating optimization algorithm was developed to improve sensing accuracy under communication and deployment constraints. Results showed that port reconfiguration adapts the array geometry to target direction, continuous refinement further improves accuracy, and range, aperture, and active port count govern sensing performance, revealing an accuracy-complexity tradeoff.

\balance
\bibliographystyle{IEEEtran}
\bibliography{refs}

@IEEEtranBSTCTL{IEEEexample:BSTcontrol,
	CTLdash_repeated_names = "no"
}

@article{Liu-2022,
	author  = {A. Liu and others},
	title   = {A survey on fundamental limits of integrated sensing and communication},
	journal = {IEEE Commun. Surveys Tuts.},
	volume  = {24},
	number  = {2},
	pages   = {994--1034},
	year    = {2022}
}

@article{Zhang-2026-ISAC,
	author  = {Z. Zhang and others},
	title   = {Finite-blocklength {ISAC} multiple access},
	journal = {IEEE Trans. Wireless Commun.},
	volume  = {25},
	pages   = {22669--22686},
	year    = {2026}
}

@inproceedings{isac_tradeoff,
	author    = {Z. Zhang and others},
	title     = {Fundamental tradeoffs for {ISAC} multiple access in finite-blocklength regime},
	booktitle = {Proc. Int. Conf. Comput. Commun. Artif. Intell. (CCAI)},
	address   = {Nanjing, China},
	pages     = {552--557},
	year      = {2026}
}

@article{Wei-2023,
	author  = {Z. Wei and others},
	title   = {Integrated sensing and communication signals toward {5G-A} and {6G}: A survey},
	journal = {IEEE Internet Things J.},
	volume  = {10},
	number  = {13},
	pages   = {11068--11092},
	month   = jul,
	year    = {2023}
}

@article{Ahmadipour-2024,
	author  = {M. Ahmadipour and others},
	title   = {An information-theoretic approach to joint sensing and communication},
	journal = {IEEE Trans. Inf. Theory},
	volume  = {70},
	number  = {2},
	pages   = {1124--1146},
	month   = feb,
	year    = {2024}
}

@article{Demirhan-2023,
	author  = {U. Demirhan and A. Alkhateeb},
	title   = {Integrated sensing and communication for {6G}: Ten key machine learning roles},
	journal = {IEEE Commun. Mag.},
	volume  = {61},
	number  = {5},
	pages   = {113--119},
	month   = may,
	year    = {2023}
}

@inproceedings{ma_isac1,
	author    = {Z. Zhang and others},
	title     = {Integrated sensing and communications for unsourced random access: A spectrum sharing compressive sensing approach},
	booktitle = {Proc. IEEE Veh. Technol. Conf. (VTC-Spring)},
	address   = {Oslo, Norway},
	pages     = {1--6},
	year      = {2025}
}

@article{ma_isac2,
	author  = {Z. Zhang and others},
	title   = {Integrated sensing and communications for unsourced random access via spectrum sharing compressive sensing approach with massive {MIMO} receiver},
	journal = {IEEE Trans. Veh. Technol.},
	volume  = {75},
	number  = {7},
	pages   = {13338--13350},
	month   = jul,
	year    = {2026}
}

@inproceedings{ma_isac3,
	author    = {Z. Zhang and others},
	title     = {Integrated sensing and communications for unsourced random access: A two-phase approach with slotted transmission},
	booktitle = {Proc. IEEE/CIC Int. Conf. Commun. China (ICCC)},
	address   = {Shanghai, China},
	pages     = {1--6},
	year      = {2025}
}

@article{massive_MIMO1,
	author  = {Y. Qi and others},
	title   = {{FBMC}-based massive connectivity with asynchronous transmission in frequency-selective fading channels},
	journal = {IEEE Trans. Wireless Commun.},
	volume  = {23},
	number  = {12},
	pages   = {18951--18968},
	month   = dec,
	year    = {2024}
}

@article{massive_MIMO2,
	author  = {Z. Zhang and others},
	title   = {Unsourced random access via random scattering with turbo probabilistic data association detector and treating collision as interference},
	journal = {IEEE Trans. Wireless Commun.},
	volume  = {23},
	number  = {12},
	pages   = {17899--17914},
	month   = dec,
	year    = {2024}
}

@article{massive_MIMO3,
	author  = {Z. Zhang and others},
	title   = {Efficient {ODMA} for unsourced random access in {MIMO} and hybrid massive {MIMO}},
	journal = {IEEE Internet Things J.},
	volume  = {11},
	number  = {23},
	pages   = {38846--38860},
	month   = dec,
	year    = {2024}
}

@article{Jiang-2023,
	author  = {H. Jiang and others},
	title   = {Hybrid far- and near-field modeling for reconfigurable intelligent surface assisted {V2V} channels: A sub-array partition based approach},
	journal = {IEEE Trans. Wireless Commun.},
	volume  = {22},
	number  = {11},
	pages   = {8290--8303},
	month   = nov,
	year    = {2023}
}

@article{Cui-2023,
	author  = {M. Cui and others},
	title   = {Near-field {MIMO} communications for {6G}: Fundamentals, challenges, potentials, and future directions},
	journal = {IEEE Commun. Mag.},
	volume  = {61},
	number  = {1},
	pages   = {40--46},
	month   = jan,
	year    = {2023}
}

@article{Liu-2023,
	author  = {Y. Liu and others},
	title   = {Near-field communications: A tutorial review},
	journal = {IEEE Open J. Commun. Soc.},
	volume  = {4},
	pages   = {1999--2049},
	month   = aug,
	year    = {2023}
}

@article{Jiang-2025NearField,
	author  = {H. Jiang and others},
	title   = {High-efficient near-field channel characteristics analysis for large-scale {MIMO} communication systems},
	journal = {IEEE Internet Things J.},
	volume  = {12},
	number  = {6},
	pages   = {7446--7458},
	month   = mar,
	year    = {2025}
}

@article{Xu2025Elastic,
  author  = {T. Xu and others},
  title   = {Elastic spectrum sensing: An adaptive
             sensing method for non-terrestrial
             communication under highly dynamic channels},
  journal = {IEEE Trans. Wireless Commun.},
  volume  = {24},
  number  = {5},
  pages   = {3868--3881},
  month   = may,
  year    = {2025},
  doi     = {10.1109/TWC.2025.3535559}
}

@article{Wong2020,
	author  = {K.-K. Wong and others},
	title   = {Fluid antenna system for {6G}: When {Bruce Lee} inspires wireless communications},
	journal = {Electron. Lett.},
	volume  = {56},
	number  = {24},
	pages   = {1288--1290},
	month   = nov,
	year    = {2020}
}

@article{New-2023fluid,
	author  = {W. K. New and others},
	title   = {Fluid antenna system: New insights on outage probability and diversity gain},
	journal = {IEEE Trans. Wireless Commun.},
	volume  = {23},
	number  = {1},
	pages   = {128--140},
	month   = jan,
	year    = {2024}
}

@article{Jiang-2025UAV,
	author  = {H. Jiang and others},
	title   = {Dynamic channel modeling of fluid antenna systems in {UAV} communications},
	journal = {IEEE Wireless Commun. Lett.},
	volume  = {14},
	number  = {10},
	pages   = {3169--3173},
	month   = oct,
	year    = {2025}
}

@article{Khamidullina-2021Bistatic,
	author  = {L. Khamidullina and others},
	title   = {Conditional and unconditional {Cram{\'e}r--Rao} bounds for near-field localization in bistatic {MIMO} radar systems},
	journal = {IEEE Trans. Signal Process.},
	volume  = {69},
	pages   = {3220--3234},
	year    = {2021}
}

@article{Wang-2024NearFieldCRB,
	author  = {H. Wang and others},
	title   = {{Cram{\'e}r--Rao} bounds for near-field sensing with extremely large-scale {MIMO}},
	journal = {IEEE Trans. Signal Process.},
	volume  = {72},
	pages   = {701--717},
	year    = {2024}
}

@article{Hua-2024NearField3D,
	author  = {H. Hua and others},
	title   = {Near-field {3D} localization via {MIMO} radar: {Cram{\'e}r--Rao} bound analysis and estimator design},
	journal = {IEEE Trans. Signal Process.},
	volume  = {72},
	pages   = {3879--3895},
	year    = {2024}
}

@article{Wang-2023NearFieldISAC,
	author  = {Z. Wang and others},
	title   = {Near-field integrated sensing and communications},
	journal = {IEEE Commun. Lett.},
	volume  = {27},
	number  = {8},
	pages   = {2048--2052},
	month   = aug,
	year    = {2023}
}

@article{Li-2024ISPAC,
	author  = {H. Li and others},
	title   = {Near-field integrated sensing, positioning, and communication: A downlink and uplink framework},
	journal = {IEEE J. Sel. Areas Commun.},
	volume  = {42},
	number  = {9},
	pages   = {2196--2212},
	month   = sep,
	year    = {2024}
}

@article{Zhang-2025DelayPhase,
	author  = {Z. Zhang and others},
	title   = {{Cram{\'e}r--Rao} bound optimization for near-field wideband {ISAC}: Delay-phase precoding for beam squint mitigation},
	journal = {IEEE Wireless Commun. Lett.},
	volume  = {14},
	number  = {11},
	pages   = {3794--3798},
	month   = nov,
	year    = {2025}
}

@article{Zhou-2025NFISACRSMA,
	author  = {J. Zhou and others},
	title   = {Flexible rate-splitting multiple access for near-field integrated sensing and communications},
	journal = {IEEE Trans. Veh. Technol.},
	volume  = {74},
	number  = {7},
	pages   = {11524--11528},
	month   = jul,
	year    = {2025}
}

@article{Luo-2025AFDM,
	author  = {Y. Luo and others},
	title   = {A novel angle-delay-{Doppler} estimation scheme for {AFDM-ISAC} system in mixed near-field and far-field scenarios},
	journal = {IEEE Internet Things J.},
	volume  = {12},
	number  = {13},
	pages   = {22669--22682},
	month   = jul,
	year    = {2025}
}

@article{Rahal-2026NearFieldISAC,
	author  = {M. Rahal and others},
	title   = {Near-field wideband {ISAC}: Estimator design and performance analysis},
	journal = {IEEE Trans. Aerosp. Electron. Syst.},
	volume  = {62},
	pages   = {16613--16627},
	year    = {2026}
}

@article{Zou-2024FASISAC,
	author  = {J. Zou and others},
	title   = {Shifting the {ISAC} trade-off with fluid antenna systems},
	journal = {IEEE Wireless Commun. Lett.},
	volume  = {13},
	number  = {12},
	pages   = {3479--3483},
	month   = dec,
	year    = {2024}
}

@article{Zhou-2024FASISAC,
	author  = {L. Zhou and others},
	title   = {Fluid antenna-assisted {ISAC} systems},
	journal = {IEEE Wireless Commun. Lett.},
	volume  = {13},
	number  = {12},
	pages   = {3533--3537},
	month   = dec,
	year    = {2024}
}

@article{Ye-2025FASSCNR,
	author  = {Y. Ye and others},
	title   = {{SCNR} maximization for {MIMO} {ISAC} assisted by fluid antenna system},
	journal = {IEEE Trans. Veh. Technol.},
	volume  = {74},
	number  = {8},
	pages   = {13272--13277},
	month   = aug,
	year    = {2025}
}

@inproceedings{Xu-2025FASCRB,
	author    = {L. Xu and others},
	title     = {Joint port selection and {Cram{\'e}r--Rao} bound optimization for {ISAC} in fluid antenna system},
	booktitle = {Proc. IEEE Int. Conf. Commun. (ICC)},
	address   = {Montreal, QC, Canada},
	pages     = {4505--4510},
	year      = {2025}
}

@article{Salem-2026FASLocalization,
	author  = {A. Salem and others},
	title   = {Fundamental limits of localization with fluid antenna systems: A {Fisher} information analysis},
	journal = {IEEE Wireless Commun. Lett.},
	volume  = {15},
	pages   = {3054--3058},
	year    = {2026}
}

@article{Zhang-2026FAA,
	author  = {Z. Zhang and others},
	title   = {Finite-aperture fluid antenna array design: Analysis and algorithm},
	journal = {IEEE Wireless Commun. Lett.},
	volume  = {15},
	pages   = {3199--3203},
	year    = {2026}
}

@article{Chen-2026NFFAS,
	author  = {S. Chen and others},
	title   = {Near-field beamforming and port selection for fluid antenna systems},
	journal = {IEEE Commun. Lett.},
	volume  = {30},
	pages   = {2248--2252},
	year    = {2026}
}

@article{Guo-2026FA,
	author  = {Y. Guo and others},
	title   = {Cramér-Rao bound optimization for fluid antenna-empowered integrated sensing and uplink communication system},
	journal = {IEEE Trans. Commun.},
	volume  = {74},
	pages   = {3631-3645},
	year    = {2026}
}

@article{Zhang-2026-aperture,
	author  = {Z. Zhang and others},
	title   = {Finite-aperture planar fluid antenna array},
	journal = {arXiv preprint},
	year    = {2026},
	url     = {https://arxiv.org/abs/2605.22040}
}

@article{fas-twc-21,
	author  = {K.-K. Wong and others},
	title   = {Fluid antenna systems},
	journal = {IEEE Trans. Wireless Commun.},
	volume  = {20},
	number  = {3},
	pages   = {1950--1962},
	month   = mar,
	year    = {2021}
}

@article{FBL3,
	author  = {Z. Zhang and others},
	title   = {Finite-blocklength fluid antenna systems},
	journal = {IEEE Trans. Wireless Commun.},
	volume  = {25},
	pages   = {22148--22164},
	year    = {2026}
}

@article{FBL4,
	author  = {Z. Zhang and others},
	title   = {Finite-blocklength fluid antenna systems with spatial block-correlation channel model},
	journal = {IEEE Wireless Commun. Lett.},
	volume  = {15},
	pages   = {1911--1915},
	year    = {2026}
}

@article{FBL1,
	author  = {Z. Zhang and others},
	title   = {On fundamental limits for fluid antenna-assisted integrated sensing and communications for unsourced random access},
	journal = {IEEE J. Sel. Areas Commun.},
	volume  = {44},
	pages   = {136--149},
	year    = {2026}
}

@article{FBL2,
	author  = {Z. Zhang and others},
	title   = {On fundamental limits of slow-fluid antenna multiple access for unsourced random access},
	journal = {IEEE Wireless Commun. Lett.},
	volume  = {14},
	number  = {11},
	pages   = {3455--3459},
	month   = nov,
	year    = {2025}
}

@article{zzt_sp1,
	author  = {Z. Zhang and others},
	title   = {Joint activity detection and channel estimation for fluid antenna system exploiting geographical and angular information},
	journal = {IEEE J. Sel. Topics Signal Process.},
	volume  = {20},
	number  = {3},
	pages   = {354--370},
	month   = apr,
	year    = {2026}
}

@article{zzt_sp2,
	author  = {Z. Zhang and others},
	title   = {Geometry-structured channel reconstruction for conventional and fluid antenna systems: {Bayesian} inference and fundamental limits},
	journal = {arXiv preprint},
	year    = {2026},
	url     = {https://arxiv.org/abs/2606.04001}
}

@article{CSI1,
	author  = {Z. Zhang and others},
	title   = {Beyond covariance: Generative spatial correlation modeling and channel interpolation for fluid antenna systems},
	journal = {arXiv preprint},
	year    = {2026},
	url     = {https://arxiv.org/abs/2604.16639}
}

@article{yuanhui_wcl,
	author  = {Y. Wu and others},
	title   = {Learned-approximate message passing under {Karhunen--Lo{\`e}ve} modeling for fluid antenna systems},
	journal = {IEEE Wireless Commun. Lett.},
	volume  = {15},
	pages   = {2719--2723},
	year    = {2026}
}

@inproceedings{GBW_zzt,
	author    = {Z. Zhang and others},
	title     = {Low-complexity {CSI} acquisition exploiting geographical diversity in fluid antenna system},
	booktitle = {Proc. IEEE Global Commun. Conf. Workshops (GC Wkshps)},
	pages     = {308--313},
	year      = {2025}
}

@article{wcsp_FAA1,
	author  = {Y. Wu and others},
	title   = {Impedance-aware zonal port activation for fluid antenna arrays},
	journal = {arXiv preprint},
	year    = {2026},
	url     = {https://arxiv.org/abs/2608.13249}
}

@article{wcsp_FAA2,
	author  = {J. Xu and others},
	title   = {Sparse port selection under mutual coupling in fluid antenna arrays},
	journal = {arXiv preprint},
	year    = {2026},
	url     = {https://arxiv.org/abs/2608.15599}
}

@article{wcsp_FAA3,
	author  = {Y. Wu and others},
	title   = {{EM}-guided graph learning for fluid antenna beamforming under current-domain constraints},
	journal = {arXiv preprint},
	year    = {2026},
	url     = {https://arxiv.org/abs/2608.13254}
}

@inproceedings{FAA_XJY,
	author    = {J. Xu and others},
	title     = {Fluid antenna-enhanced flexible beamforming},
	booktitle = {Proc. IEEE Wireless Commun. Netw. Conf. Workshops (WCNCW)},
	pages     = {1--6},
	year      = {2026}
}

@article{wcsp_FAA4,
	author  = {H. Liang and others},
	title   = {Peak sidelobe suppression in planar fluid antenna array},
	journal = {arXiv preprint},
	year    = {2026},
	url     = {https://arxiv.org/abs/2606.31149}
}

@article{EM-FAA,
	author  = {Z. Zhang and others},
	title   = {Electromagnetic-aware fluid antenna array},
	journal = {arXiv preprint},
	year    = {2026},
	url     = {https://arxiv.org/abs/2607.21375}
}

@inproceedings{FAS_channel_0_2,
	author    = {K. Zhou and others},
	title     = {Sparsity-exploiting channel estimation for unsourced random access with fluid antenna},
	booktitle = {Proc. IEEE Veh. Technol. Conf. (VTC-Spring)},
	address   = {Oslo, Norway},
	pages     = {1--7},
	year      = {2025}
}

@inproceedings{ML_WCNC,
	author    = {H. Liang and others},
	title     = {Neural networks-enabled channel reconstruction for fluid antenna systems: A data-driven approach},
	booktitle = {Proc. IEEE Wireless Commun. Netw. Conf. Workshops (WCNCW)},
	pages     = {1--6},
	year      = {2026}
}

\end{document}